\RequirePackage{silence}
\documentclass[12pt,a4paper]{article}
\usepackage{jheppub}%
\usepackage{Preamble}%
\usepackage{Settings}%

\begin{document}
\title{Neutrinoless double-beta decay at colliders: interference between Majorana states}

\author{Jonathan L.\ Schubert, Oleg Ruchayskiy}
\emailAdd{Oleg.Ruchayskiy@nbi.ku.dk} \affiliation{Niels Bohr Institute,
  University of Copenhagen, Blegdamsvej 17, DK-2010, Copenhagen, Denmark}

\abstract{Heavy neutral leptons (HNLs) are hypothetical particles able to explain several puzzles of fundamental physics, first and foremost --- neutrino oscillations.
Being \emph{sterile} with respect to Standard Model interactions, these particles admit Majorana masses, allowing for violation of the total lepton number. Lepton number violating (LNV) processes thus become a key signature of HNLs, pursued by many experiments.
In this work we demonstrate that if HNLs are the sole origin of neutrino masses, destructive interference between Majorana states suppresses the same-sign di-lepton signal.
In the phenomenologically interesting case of large HNL couplings, such a suppression is akin to the cancellation of HNLs' contributions to neutrino masses.
Nevertheless, the signal can be much larger than coming from the Weinberg operator alone.
We identify regions of the parameter space of such realistic HNL
models where the LNV signal is maximised at the LHC and
future FCC-hh.
Our results are obtained within the effective W approximation which allows for
analytic treatment and gives a clear dependence on the model parameters.  Although
approximate, they are argued to be correct within a factor of few.
}
\maketitle
\flushbottom

\section{Introduction}
\label{sec:introduction}

Searches for processes forbidden by the symmetries of the Standard Model provide promising avenue of probing for new physics.
Lepton Number Violating processes (LNV) have always been among prime candidates for such searches (see e.g.\cite{Goeppert-Mayer:1935uil,Pontecorvo:1957vz,Pontecorvo:1968wp,Weinberg:1979sa,Wilczek:1979hc,Lazarides:1980nt,Mohapatra:1986bd}).
The motivation for it comes from the century-old observation of Ettore Majorana \cite{Majorana:1937vz} that neutral particles admit a different type of mass, \emph{Majorana mass}.
A main experimental approach to LNV searches is the neutrinoless double-beta decay (\nubb, see \cite{DellOro:2016tmg,Cirigliano:2022oqy} and refs.\ therein).
Other LNV probes include experiments with muonic atoms \cite{SINDRUMII:1998mwd} and LNV decays of mesons and $\tau$-leptons (see e.g.\ \cite{Belle:2012unr,NA62:2019eax,BaBar:2012eip,LHCb:2013hxr}).
Collider searches, such as \ppll \cite{ATLAS:2011izm,ATLAS:2012ak,CMS:2012wqj,ATLAS:2015gtp,CMS:2015qur,CMS:2016aro,ATLAS:2018dcj,CMS:2018agk,CMS:2018jxx,CMS:2018iye,LHCb:2020wxx}, provide a powerful tool to probe for LNV, especially in the sector involving particles other than electrons.
The appeal of LNV searches lies in their potential to deduce the presence of more massive particles than can be accessed directly.


In this work we consider the \emph{\typeI model} \cite{Minkowski:1977sc,Yanagida:1979as,Glashow:1979nm,Gell-Mann:1979vob,Mohapatra:1979ia,Mohapatra:1980yp,Schechter:1980gr,Schechter:1981cv}
in which the lepton number violation comes from \emph{heavy neutral leptons} (HNLs).
The parameters of these HNLs are chosen so as to generate (Majorana) masses of neutrinos and explain neutrino oscillations.
The \typeI model contains two sets of Majorana particles: HNLs with masses $m_{\Nu_I}$ anywhere from $\si{eV}$ to $10^{15}\,\si{GeV}$; and mass states of active neutrinos with $m_{\nu_i}$ being in sub-eV range as dictated by neutrino oscillations~\cite{Esteban:2020cvm} in combination with cosmological measurements (see \cite{Dvorkin:2019jgs} and references therein).
HNLs interact with $W^\pm$, $Z^0$ and the Higgs boson similarly to the Standard Model neutrinos but with the couplings suppressed by the flavour-dependent mixing angles $\mix_{\alpha I}$.\footnote{Our notations: flavour index: $\alpha = \{e,\mu,\tau\}$; HNLs are enumerated with the index $I=1,\dots,\mathcal{N}$.
  HNL masses: $m_{\Nu_I}$; the elements of the $3\times \mathcal N$ matrix $\mix_{\alpha I}$ describes interactions of $I^{th}$ HNL with the flavour $\alpha$.}

\begin{figure}[!t]
  \centering
  \begin{minipage}{0.445\linewidth}
    \centering
    \subfloat[t-channel diagram][WBF process, reminiscent of \nubb
    decay,  used to interpret same-sign lepton searches in \cite{CMS:2022rqc}. ]{
      \label{fig:main_process}
\begin{tikzpicture}
  \begin{feynman}

    \vertex (i);
    \vertex [right =of i] (qcd1);
    \vertex [left =of qcd1] (ui11){\(u\)};
    \path (qcd1) ++ (1,0.33) node[vertex] (dummy1);
    \vertex [right =of dummy1] (df11){\(d\)};
    \path (qcd1) ++ (0,0.33) node[vertex] (qcd11);
    \path (qcd1) ++ (0,0.66) node[vertex] (qcd12);
    \vertex [left =of qcd11] (ui12) {\(u\)};
    \vertex [left =of qcd12] (di1) {\(d\)};

    \vertex [above right =of qcd11] (jet1dummy1);
    \vertex [above right =of qcd12] (jet1dummy2);
    \vertex [right =of jet1dummy1] (uf1){\(u\)};
    \vertex [right =of jet1dummy2] (df12){\(d\)};

    \vertex [below right =of qcd1] (ew1);
    \vertex [right =of ew1] (fl1){\(\ell^{+}\)};
    \vertex [below =of ew1] (ew2);
    \vertex [right =of ew2] (fl2){\(\ell^{+}\)};

    \vertex [below left =of ew2] (qcd2);
    \vertex [left =of qcd2] (ui21){\(u\)};
    \path (qcd2) ++ (1,-0.33) node[vertex] (dummy2);
    \vertex [right = of dummy2] (df21){\(d\)};
    \path (qcd2) ++ (0,-0.33) node[vertex] (qcd21);
    \path (qcd2) ++ (0,-0.66) node[vertex] (qcd22);
    \vertex [left =of qcd21] (ui22) {\(u\)};
    \vertex [left =of qcd22] (di2) {\(d\)};

    \vertex [below right =of qcd21] (jet2dummy1);
    \vertex [below right =of qcd22] (jet2dummy2);
    \vertex [right =of jet2dummy1] (uf2){\(u\)};
    \vertex [right =of jet2dummy2] (df22){\(d\)};

    \path (qcd1) ++ (-.5,-0.33) node[vertex] (RectUL);
    \path (qcd2) ++ (-.5,+0.33) node[vertex] (RectLL);
    \path (RectUL) ++ (+4,0) node[vertex] (RectUR);
    \path (RectLL) ++ (+4,0) node[vertex] (RectLR);
    \diagram* {

      (ui11) -- [fermion, black!40!cyan, thick] (qcd1) -- [fermion, black!40!magenta, thick] (df11),
      (ui12) -- [fermion, black!40!cyan, thick] (qcd11) -- [fermion, black!40!cyan, thick] (uf1),
      (di1) -- [fermion, black!40!magenta, thick] (qcd12) -- [fermion, black!40!magenta, thick] (df12),

      (qcd1) -- [boson, edge label'=\(W\), black!40!blue, thick] (ew1),
      (fl1) -- [fermion, orange, thick] (ew1),
      (qcd2) -- [boson, edge label'=\(W\), black!40!blue, swap, thick] (ew2),
      (fl2) -- [fermion, orange, thick] (ew2),
      (ew1) -- [plain, edge label'=\(\mathrm{n}\), black!60!green, very thick, swap] (ew2),

      (ui21) -- [fermion, black!40!cyan, thick] (qcd2) -- [fermion, black!40!magenta, thick] (df21),
      (ui22) -- [fermion, black!40!cyan, thick] (qcd21) -- [fermion, black!40!cyan, thick] (uf2),
      (di2) -- [fermion, black!40!magenta, thick] (qcd22) -- [fermion, black!40!magenta, thick] (df22),

    };
    \draw (ew1) node[dot, fill, draw=black] {};
    \draw (ew2) node[dot, fill, draw=black] {};
    \draw (qcd1) node[dot, fill, draw=black] {};
    \draw (qcd2) node[dot, fill, draw=black] {};

  \end{feynman}
\end{tikzpicture} }
  \end{minipage}\hfill
  \begin{minipage}{0.545\linewidth}
    \centering
    \subfloat[s-channel diagram][Drell-Yan process, used to interpret same-sign lepton searches in \cite{ATLAS:2011izm,ATLAS:2012ak,CMS:2012wqj,ATLAS:2015gtp,CMS:2015qur,CMS:2016aro,ATLAS:2018dcj,CMS:2018agk,CMS:2018jxx,CMS:2018iye}.]{%
      \label{fig:s-channel_lljj}
\begin{tikzpicture}
  \begin{feynman}
    \vertex (i);
    \vertex [below =of i] (idummy);
    \vertex [right =of idummy] (v1);
    \vertex [above left =of v1] (i1) {\(q\)};
    \vertex [below left =of v1] (i2) {\(\overline{q^\prime}\)};
    \vertex [right =of v1] (v2);
    \vertex [above right =of v2] (f1) {\(\ell^+\)};
    \vertex [below right =of v2] (v3);
    \vertex [above right =of v3] (f2) {\(\ell^+\)};
    \vertex [below right =of v3] (v4);
    \vertex [above right =of v4] (j1) {\(q\)};
    \vertex [below right =of v4] (j2) {\(\overline{q^\prime}\)};

    \diagram* {
      (i1) -- [fermion, black!40!cyan, thick] (v1) -- [fermion, black!40!magenta, thick] (i2),
      (v1) -- [boson, edge label'=\(W\), black!40!blue, thick] (v2),
      (f1) -- [fermion, orange, thick] (v2),
      (v2) -- [plain, edge label'=\(\mathrm{n}\), black!60!green, very thick] (v3),
      (f2) -- [fermion, orange, thick] (v3),
      (v3) -- [boson, edge label'=\(W\), black!40!blue, thick] (v4),
      (j2) -- [fermion, black!40!cyan, thick] (v4) -- [fermion, black!40!magenta, thick] (j1),
    };

    \draw (v1)  node[dot, fill, draw=black] {};
    \draw (v2) node[dot, fill, draw=black] {};
    \draw (v3) node[dot, fill, draw=black] {};
    \draw (v4) node[dot, fill, draw=black] {};

  \end{feynman}
\end{tikzpicture}}
  \end{minipage}
  \caption[LHC searches for lljj]{Two kinds of processes leading to same-sign
    leptons plus jets signatures in $pp$ collisions.
    Additional diagrams, originating from the interchange of the final states are not shown.
    The symbol $\mathrm{n}$ represents both light and heavy Majorana states.
    Panel \protect\subref{fig:main_process}: Majorana particles in $t$-channel ($W$ boson fusion process).
    Panel \protect\subref{fig:s-channel_lljj}: Majorana particles in $s$-channel (Drell-Yan process).
    The contribution from two processes are incoherent because of the antiquarks in the final state of process~\subref{fig:s-channel_lljj} as compared to \subref{fig:main_process}.
  }
  \label{fig:lljj}
\end{figure}
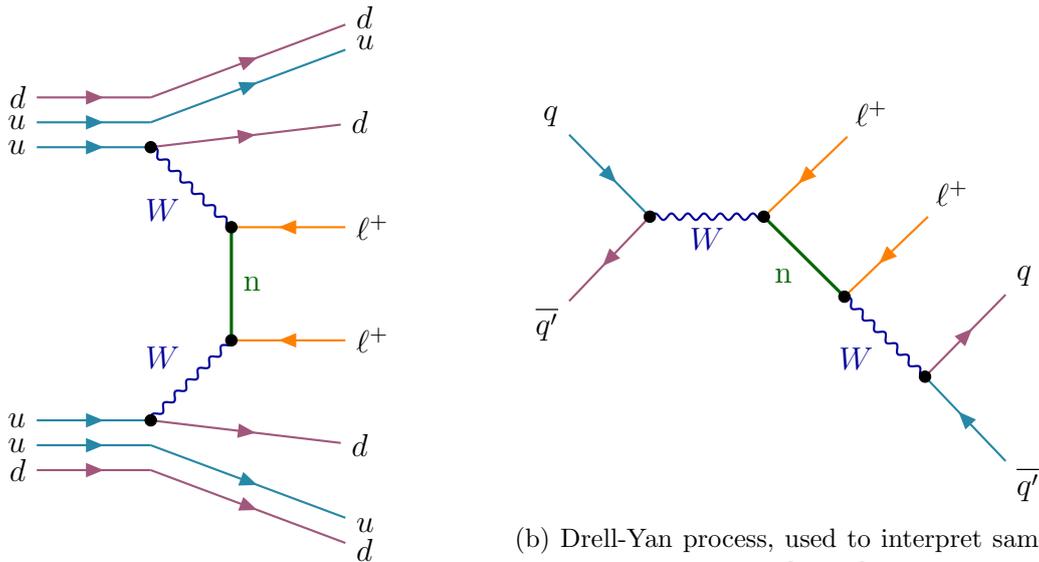

Many theoretical proposals on how to probe LNV signatures of HNLs have been put forward (see e.g.\ \cite{Dicus:1991fk,Datta:1993nm,Ali:2001gsa,Panella:2001wq,deGouvea:2007qla,delAguila:2005pin,delAguila:2006bda,Han:2006ip,delAguila:2007ua,Chen:2008qb,FileviezPerez:2008wbg,Chao:2009ef,Frandsen:2009fs,Altarelli:2010gt,Ibarra:2011xn,Han:2012vk,Angel:2012ug,Helo:2013ika,Helo:2013dla,deGouvea:2013zba,King:2014nza,Deppisch:2015qwa,Ng:2015hba,Das:2016hof,Anamiati:2016uxp,Das:2017nvm,Abada:2017jjx,Cai:2017mow,Chun:2019nwi,Drewes:2019byd,Tastet:2019nqj,DeVries:2020jbs,Gargalionis:2020xvt,deGouvea:2021rpa,Zhou:2021lnl,Aoki:2020til,Fuks:2020att,Fuks:2020zbm,Zapata:2022qwo} for an incomplete list of references).
In this work we will concentrate on a single signature of HNLs at hadron colliders: the process with two same-sign leptons and two jets (and no missing energy) in the final state, \ppll.
The corresponding process may be mediated either via HNLs in the $s$-channel (Drell-Yan process, \cref{fig:s-channel_lljj} \cite{delAguila:2007qnc,Degrande:2016aje}) or via the direct analog of \nubb with HNLs in $t$-channel --- \emph{$W$ boson fusion} (WBF), \cref{fig:main_process}, \cite{Datta:1993nm,Ali:2001gsa,Panella:2001wq,Chen:2008qb,Fuks:2020att}.
Most of the previous LHC searches \cite{ATLAS:2011izm,ATLAS:2012ak,CMS:2012wqj,ATLAS:2015gtp,CMS:2015qur,CMS:2016aro,ATLAS:2018dcj,CMS:2018agk,CMS:2018jxx,CMS:2018iye,LHCb:2020wxx} concentrated on the process shown in \cref{fig:s-channel_lljj}.
The contribution of a single HNL to the WBF process and the corresponding Monte Carlo level analysis of the LHC sensitivity to such a signal were considered recently in \cite{Fuks:2020att}.
It was found that the LHC sensitivity to these types of searches may be more prominent than the Drell-Yan process for HNLs with masses $m_\Nu \gtrsim\SI{1}{TeV}$.
The corresponding analysis was performed recently by the CMS collaboration~\cite{CMS:2022rqc}.

However, in realistic HNL models, responsible for neutrino masses, the situation is more complicated.
Indeed, experimentally two mass splittings in the light neutrino mass states have been measured~\cite{Esteban:2020cvm}.
Therefore, \emph{at least two} HNLs should be considered in the framework of the \typeI model to explain neutrino data.
This means that even the simplest realistic model contains \emph{four} Majorana particles: two light (active) neutrinos and two HNLs.\footnote{To avoid potential confusion, we remind that in the \typeI model, two non-degenerate Majorana states are sufficient to account for two light neutrino mass states. In \emph{Inverse}
  \cite{Mohapatra:1986aw, Mohapatra:1986bd, Bernabeu:1987gr} or \emph{Linear}
  \cite{Akhmedov:1995ip, Akhmedov:1995vm} Seesaw models at least four
   HNLs with opposite CP phases are required.}
All these four Majorana particles contribute coherently to the same LNV process and their interference may be important.

An example of such an interplay is known from the studies of the similar process -- neutrino-less double beta decay (\nubb) \cite{Cirigliano:2022oqy,DellOro:2016tmg}.
On the one hand, the contributions to the process of light and heavy states generically interfere destructively such that the resulting signal, expressed via \emph{effective neutrino mass} $m_{\beta\beta}$ is smaller than the one, provided by the light Majorana neutrinos alone, $m_{\beta\beta}^\nu$.\footnote{The literature on the subject is vast, see e.g.\  \cite{Halprin:1983ez,Benes:2005hn,Bezrukov:2005mx,Blennow:2010th,Mitra:2011qr,Lopez-Pavon:2012yda,Asaka:2013jfa,Faessler:2014kka,Lopez-Pavon:2015cga,Hernandez:2016kel,Drewes:2016lqo,Asaka:2020lsx,Asaka:2020wfo,Asaka:2021hkg}.}
On the other hand, enhancements of the signal as compared with $m_{\beta\beta}^\nu$ is possible for some values of HNL parameters \cite{Blennow:2010th,Ibarra:2010xw,Ibarra:2011xn,Mitra:2011qr,Lopez-Pavon:2012yda, Lopez-Pavon:2015cga,Hernandez:2016kel,Drewes:2016lqo}, especially if the spectrum of HNLs is hierarchical \cite{Asaka:2020lsx,Asaka:2020wfo,Asaka:2021hkg}.\footnote{This, however, may require fine-tuning between tree level and one-loop contributions to neutrino masses, see below.}
The question, therefore, arises --- \emph{how do light and heavy Majorana degrees of freedom interplay in case of the WBF process at colliders}?
Does cancellation occur between the states?
Can the signal get enhanced for some values of HNL parameters?
This paper is devoted to answering these questions.

\paragraph{Plan of the paper:}

The paper is organised as follows.
In \cref{sec:introduction-to-analysis} we argue that in order to estimate the contributions of Majorana particles to the $pp\to \ell^\pm\ell^\pm jj$ cross section, it is sufficient to consider same-sign $WW$ scattering to leptons.
This makes our treatment analytic and allows to explore the dependence of the cross section on the parameters of the model.
We then show that if HNLs are responsible for neutrino masses, the contributions of several HNLs can cancel each other (\cref{appendix:1-hnl-case} further substantiates this statement).
Based on the analysis of \cref{sec:introduction-to-analysis} we can anticipate in what parts of the parameter space the cross section can nevertheless become sizeable.

In \cref{sec:WW_ll_2HNLs} we compute same-sign $WW$ scattering in a model with two HNLs.
We show that, as expected, quasi-degenerate HNLs give negligible contribution to LNV processes, while for hierarchical HNLs the contribution can become sizeable.
We substantiate the simplified treatment of \cref{sec:WW_ll_2HNLs} by exact computations in \cref{appendix:CasasIbarra}.

Our results are summarised in \cref{sec:pp_lljj}.
They demonstrate that the total cross section for the process $pp \to \ell^\pm \ell^\pm jj$ is at best  $\mathcal{O}(\SI{1}{fb})$ at $\sqrt{s_\lhc} \simeq \SI{13}{TeV}$, making its exploration challenging even during the high-luminosity LHC (HL-LHC) phase.
Owing to the higher centre of mass energy --- and resultantly larger  $pp \to \ell^\pm \ell^\pm jj$ cross section --- we find that the proposed Future Circular Collider, during its $\sqrt{s_\fcc} \simeq \SI{100}{TeV}$ hadron-hadron phase (FCC-hh), could produce competitive bounds in the $m_\Nu\sim\few\,\si{TeV}$ regime.
Both these results require the HNL masses to differ by a factor of few.

We also discuss  (\cref{sec:equiv-betw-lnv}) whether, as commonly believed, LNV signals in the type-I seesaw model are proportional to neutrino masses.
\Cref{appx:type-I-seesaw,,appendix:1-hnl-case,,appendix:CasasIbarra,,appx:effectiveWApprox} contain additional computations as well as basic definitions, needed for completeness of our results.
In \cref{appx:FutureColliderNuBB} we present possible points of departure for future works.

\section{Main idea and results}
\label{sec:introduction-to-analysis}

\begin{figure}[!t]
  \centering 

\tikzset{cross/.style={cross out, draw=black, minimum size=2*(#1-\pgflinewidth), inner sep=0pt, outer sep=0pt},
cross/.default={8pt}}
\begin{tikzpicture}
  \begin{feynman}
    \vertex (i1){\(W^{+}\)};
    \vertex [below =of i1] (fill1);
    \vertex [below =of fill1] (fill2);
    \vertex [below =of fill2] (i2) {\(W^{+}\)};

    \vertex [right =of i1] (v1);
    \vertex [below =of v1] (o1);
    \vertex [below =of o1] (o2);
    \vertex [right =of i2] (v2);
    \vertex [right =of v1] (f1) {\(\ell^{+}\)};
    \vertex [right =of v2] (f2) {\(\ell^{+}\)};

    \diagram* {
      (i1) -- [boson, black!40!blue, thick] (v1),
      (f1) -- [fermion, orange, thick, momentum={[arrow style=black]\(k_1\)}] (v1),
      (i2) -- [boson, black!40!blue, thick] (v2),
      (f2) -- [fermion, orange, thick, momentum={[arrow style=black, yshift=6mm ,swap]\(k_2\)}] (v2),
      (v1) -- [fermion, orange, thick, edge label'=\(\nu_L\)] (o1),
      (v2) -- [fermion, orange, thick, edge label'=\(\nu_L\), swap] (o2),
      (o1) -- [plain, edge label'=\(\nu_R\), black!60!green, very thick, momentum={[arrow style=black!20!red,distance=10mm,]\(p\)}] (o2),
    };
    \draw (o1) node[cross out, draw=black,thick] {};
    \draw (o2) node[cross out, draw=black,thick] {};
    \draw (v1) node[dot, fill, draw=black] {};
    \draw (v2) node[dot, fill, draw=black] {};


  \end{feynman}
\end{tikzpicture}
  \caption{Feynman Diagram of the two same sign $W$-bosons to two same sign
    leptons in the flavour basis to first order in $\mix^2$.  The crosses mark a
    transition from standard model neutrino to right handed sterile neutrino,
    where the coupling is the Dirac mass $(m_D)_{I\ell}$.}
  \label{fig:FeynDiag WWll Scattering Flavour}
\end{figure}
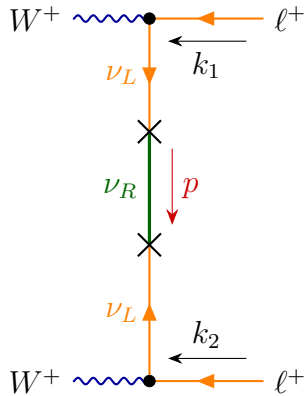

The process in \cref{fig:main_process} receives contributions from all Majorana particles (HNLs and light neutrinos).
An analysis of their relative contribution to the LNV processes and potential interference can be performed at the level of a $2\to 2$ process
\begin{equation}
  \label{eq:WW_ll}
  W^\pm W^\pm \to \ell_\alpha^\pm \ell_\beta^\pm
\end{equation}
(where $\ell_{\alpha,\beta}^\pm$ are any of the Standard Model charged leptons, not necessarily of the same flavour).
Once the cross section for the process in \cref{eq:WW_ll} is known, one can approximately reconstruct the full $pp \to \ell_\alpha^\pm \ell_\beta^\pm jj$ cross section by using the \emph{effective W-boson
approximation} (or EWA)\footnote{The PDFs of this approximation are derived at leading QCD order, but includes further approximations (for discussion see \cref{appx:effectiveWApprox}). Monte Carlo studies \cite{Fuks:2020att} demonstrate that the $k$-factor (NLO vs.\ LO correction) is $\lesssim 1.5$ for the processes in question, supporting this approach.}~\cite{Dawson:1984gx,Kane:1984bb,Kunszt:1987tk}:
\begin{multline}
  \label{eq:eff_W_approx}
  \sigma(\ppll) = \\
  2\sum_{%
    \begin{subarray}{c}
      \lambda,\lambda^\prime=\{\mathrm{T,L}\}   \\
      \xi = \pm\
    \end{subarray}
  }\int \dd{x_1}\dd{x_2} f_{W^\xi}^\lambda(x_1) f_{W^\xi}^{\lambda^\prime}(x_2) \sigma^{\lambda\lambda^\prime}(W^\xi W^\xi \to \ell^\xi \ell^\xi)(x_1x_2s_{pp}).
\end{multline}
where ${x}_1,{x}_2$ are the $W$'s Bj{\"o}rken $x$, $s_{pp}$ is the $pp$ centre of mass energy, and the sum goes over both signs of $W$-bosons ($\xi$)
as well as the transversal (T) and longitudinal (L) polarizations.

In this approximation, $W$-bosons are treated as polarised partons, and the $pp$ level cross section is derived by folding the polarised cross sections of the $WW$ scattering process, $\sigma^{\lambda\lambda^\prime}(W^\pm W^\pm \to \ell^\pm \ell^\pm)$, with the corresponding parton distribution functions (PDFs)~$f_{W^\pm}^\lambda(x)$.
More details are provided in the Appendix~\ref{appx:effectiveWApprox}.
The caveats and limitations of this approximation have been recently analysed in \cite{Ruiz:2021tdt} (see also \cite{Kunszt:1987tk,Ruiz:2015gsa}).

\begin{figure}[!t]
  \centering
    \label{fig:ExclusionLimitResults}
  \includegraphics[width=\linewidth]{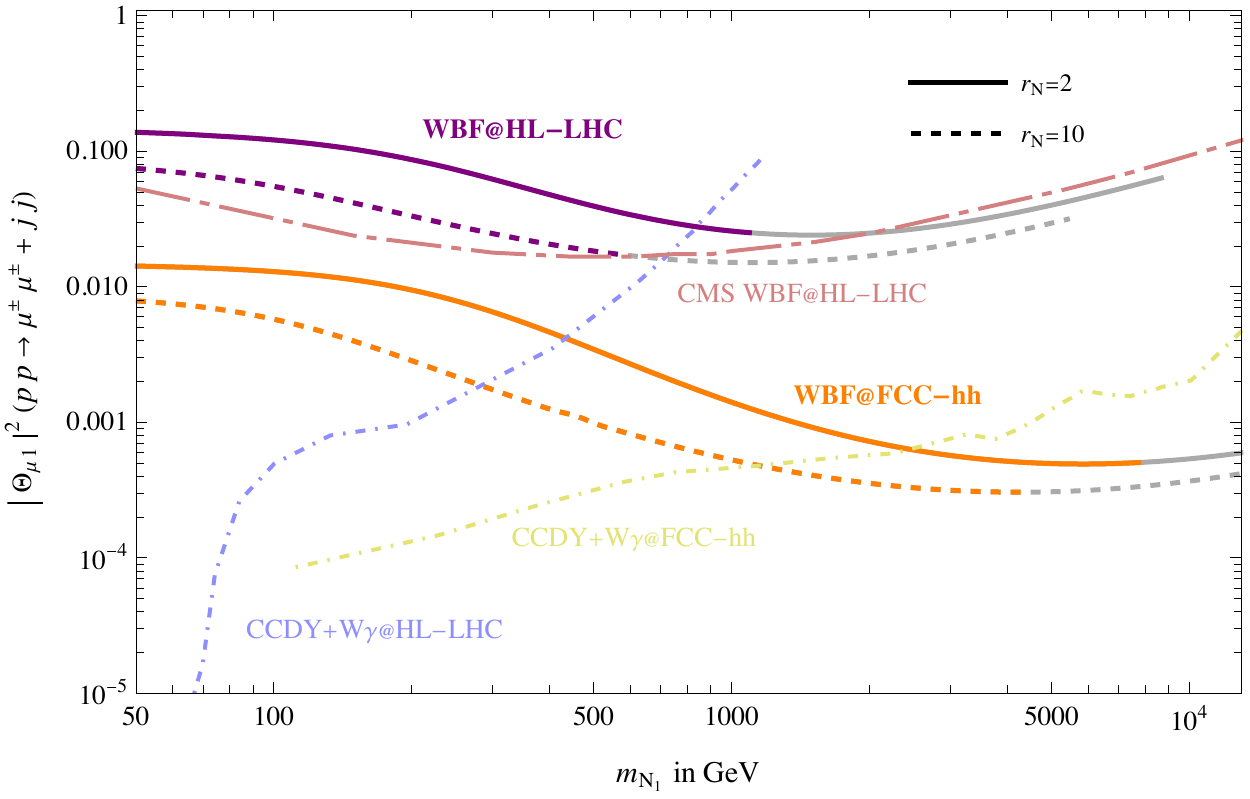}
  \caption[Exclusion limits on the HNL mixing angle from the WBF
  processes.]{Exclusion limits on the HNL mixing angle
    for WBF-mediated process $pp\to\mu^\pm\mu^\pm+jj$ at the HL-LHC and FCC-hh (purple and orange lines).
    The results are presented for two HNLs whose mass ratio equals to $r_\Nu$, with $\mix_{\ell 1}$ being the mixing angle of the lighter HNL (see in-plot legends).
    The long-short dashed line indicates the rescaling of the recent CMS
    dilepton search  \cite{CMS:2022rqc} to HL-LHC phase (red-grey).
    Short dash-dotted ``$\mathrm{CCDY}+W\gamma$'' lines are  extrapolations of the CMS tri-lepton searches \cite{CMS:2018CCDY} based on \cite{Alva:2014gxa} (light blue) and a \madgraph simulation based on the ATLAS detector \cite{Pascoli:2018heg} (yellow-grey).
    All limits are at $\SI{95}{\%}$ \textsc{c.l.}
    Grey ends of the orange and purple curves correspond to Yukawa
    couplings approaching non-perturbative values   $1<|F_{\ell
      1}|<4\pi$. See text for details.}
\end{figure}

Going fast-forward, the results of our analysis are shown in \cref{fig:ExclusionLimitResults} in terms of parameter exclusion on the HNL mixing angle $\mix_{\ell1}$ at $\SI{95}{\%}$ \textsc{c.l.}
without detector cuts and assuming background-free searches for the high-luminosity LHC phase and FCC-hh.
The details and assumptions that went into these exclusion limits are presented in \cref{sec:WW_ll_2HNLs,,sec:pp_lljj} below.
For a qulaititative discussion of these results see \cref{sec:discussion}.

The qualitative analysis of the matrix element for \cref{eq:WW_ll} is most readily performed in the \emph{flavour basis}.
In this case only the right-chiral neutrinos $\nu_R$ contain a Majorana mass term.\footnote{\Cref{sec:WW_ll_2HNLs} below substantiates these arguments by proper computation in the mass basis.}
Therefore, analysis of the diagram in \cref{fig:FeynDiag WWll Scattering Flavour} includes all contributions and interference terms.

The amplitude of the $\WWll$ subprocess shown in \cref{fig:FeynDiag WWll Scattering Flavour} is written as
\begin{equation}
  \label{eq:WW Flavour Amplitude}
  \mathcal{M}_\ww^{\mu\nu} = -i\frac{g^2}{2}\bar u(k_2) \gamma^\mu P_R \frac1{\slashed p} \sum_I\left[(m_D)_{\alpha I}\frac{\slashed{p} + m_{\Nu_I}}{p^2 - m_{\Nu_I}^2} (m_D)_{\beta I}\right]\frac1{\slashed p}\gamma^\nu P_L v(k_1) + (k_1 \leftrightarrow k_2).
\end{equation}
here $g$ is the weak coupling constant, $P_{R,L}$ are chiral projectors, $u,v$ are spinors of the outgoing leptons. Dirac ($m_D$) and Majorana ($m_{\Nu_I}$) masses are defined in \cref{appx:type-I-seesaw}.
We have implicitly chosen a counterclockwise evaluation direction, the
effects of which --- according to the Feynman rules for fermion number violating interactions
\cite{Denner:1992vza} --- are twofold:
\begin{compactitem}[--]
\item The spinor associated with the fermion going against the chosen
  direction is charge conjugated.
\item The chiral projector in the vertex connecting to this fermion is
  transformed according to $\Gamma^\prime = C\Gamma^TC^{-1}$, where $C$ is the charge-conjugation matrix.
\end{compactitem}
These rules result in the cancellation of the $\slashed{p}$ term in the numerator
of \cref{eq:WW Flavour Amplitude}, so that the amplitude is
proportional to $m_{\Nu_I}$. This cancellation can be anticipated from the fact that LNV effects are
enabled solely by the existence of the Majorana mass.  We can rewrite
\cref{eq:WW Flavour Amplitude} in the compact form
\begin{equation}
  \label{eq:Compact WW Flavour Amplitude}
  \mathcal{M}_{\ww}^{\mu\nu} = -i\frac{g^2}{2}\biggl(\bar u(k_1) \gamma^\mu \gamma^\nu P_L v(k_2)\biggr) \times \sum_I \mix_{\alpha I}\mix_{\beta I}\left[\frac{m_{\Nu_I}^3}{t(t - m_{\Nu_I}^2)} - (t\leftrightarrow u)\right],
\end{equation}
where $t$ and $u$ are Mandelstam variables.

At this stage we can already anticipate an answer to the questions raised in \cref{sec:introduction}:
In the limit $|t|,|u| \ll m_{\Nu_I}^2$, the sum in \cref{eq:Compact WW Flavour Amplitude} reduces to the seesaw expression (see \cref{appx:type-I-seesaw}):
\begin{equation}
  \label{eq:6}
  m_{\alpha\beta}^{(\nu)} \equiv \sum_{i=1}^3 V_{\alpha i} m_i V_{\beta i} = -\sum_I \mix_{\alpha I}\mix_{\beta I} m_{\Nu_I}.
\end{equation}
where $V_{\alpha i}$ are PMNS matrix elements and $m_i$ are light neutrino masses.
This means that for very heavy HNLs the total cross section will be proportional to the neutrino masses, i.e.\ will be unobservable.

Similarly, for \emph{almost degenerate} HNLs ($m_{\Nu_1}\simeq m_{\Nu_2} = m_{\Nu}$) \cref{eq:Compact WW Flavour Amplitude} becomes proportional to a linear combination of $m_{\Nu}\sum_I \mix_{\alpha I} \mix_{\beta I}$ which again reduces to the light neutrino mass matrix~(\ref{eq:6}).
This result is not surprising, as for two quasi-degenerate HNLs the total lepton number becomes conserved~\cite{Shaposhnikov:2006nn,Kersten:2007vk}.

However, in the case where $|t|,|u| \sim m_{\Nu_I}^2$ and/or HNLs are non-degenerate, \cref{eq:Compact WW Flavour Amplitude} does not readily reduce to the seesaw relation and thus is not proportional to the neutrino masses.
Hence, \emph{one can expect that the largest LNV effect can be achieved for the HNLs with hierarchical masses.}
The following sections substantiate this claim.

\section[Same-sign W scattering]{$W^\pm W^\pm\to\ell^\pm\ell^\pm$
  scattering in the model with two HNLs}
\label{sec:WW_ll_2HNLs}

HNLs with mass $m_{\Nu_I}$ and coupling $\mix_{\ell I}$ contribute to neutrino mass matrix at the level $\mix_{\ell I}^2 m_{\Nu_I}$.
Therefore, to be consistent with neutrino data, one has two distinct possibilities:
either mixings $\mix_{\ell I}$ are close to the \emph{seesaw line} (see \Cref{appx:type-I-seesaw}):
\begin{equation}
  \label{eq:seesaw_mixing}
  |\mix_{\ell I}|^2 \sim 5\times 10^{-14}\left(\frac{\si{TeV}}{m_\Nu}\right)\;,
\end{equation}
or several HNLs cancel each other's contributions to neutrino masses~\cite{Shaposhnikov:2006nn,Kersten:2007vk}.
This latter case is phenomenologically interesting, as HNL mixings can exceed the naive limit~\eqref{eq:seesaw_mixing} by orders of magnitude, making them accessible for direct searches.
We will, therefore, concentrate on this case.

\subsection[Quasi-Dirac-like model]{Quasi-Dirac-like model -- two HNLs with cancelling contribution to neutrino mass}
\label{sec:qdl}

To simplify our treatment, we arrange the HNL mixing angles in such a way that the individual contributions to the light neutrino masses cancel:\footnote{In this Section we ignore the flavour structure and consider a generic flavour $\ell$.}
\begin{equation}
  \label{eq:toy_2HNL_model}
  0=m_{\ell\ell}^{\nu}= \sum_{I=1}^2 \mix_{\ell I}^2m_{\Nu_I} .
\end{equation}
This allows getting rid of light neutrino contributions and analysing only interference between HNL states.
We will see below that the contribution to the WBF process will be non-zero in this case, because for large HNL mixings, neutrino contributions are subdominant.
The effect from light neutrino masses is presented in \cref{appendix:CasasIbarra}.
Finally, this parameter choice also allows us to demonstrate that LNV can be
prominent, even if neutrino masses are zero.

In the context of 2 HNLs, \cref{eq:toy_2HNL_model} can be satisfied by the prescription of
\begin{equation} \label{Eq. Masslessness Condition}
  \mix_{\ell2}=\pm i\mix_{\ell1}\sqrt{\frac{m_{\Nu_1}}{m_{\Nu_2}}}=\pm i\frac{\mix_{\ell1}}{\sqrt{r_\Nu}},
\end{equation}
where we have introduced the \emph{mass ratio}
\begin{equation}
  \label{eq:rN}
  r_\Nu\equiv\tfrac{m_{\Nu 2}}{m_{\Nu 1}} \ge 1.
\end{equation}
We will refer to the HNL pair obeying \cref{Eq.
  Masslessness Condition} with $r_\Nu > 1$ as the \emph{quasi-Dirac-like} (\qdl) model.

In the limit of $r_\Nu\to 1$, known as the \emph{Dirac-limit}, the two HNLs will form a single Dirac fermion.
As a consequence, SM lepton number is conserved and the LNV effects will cancel entirely.
This is not the case if $r_\Nu > 1$ where, as we will see, LNV effects are non-zero.
The \qdl model produces realistic predictions in the large mixing angle regime, which we will present in the following.
By comparison, the effect of light neutrino masses is marginal (presented in \Cref{appendix:CasasIbarra}).

\subsection[Cross section of the 2-to-2 process]{\boldmath Cross section of the \(W^\pm W^\pm\to\ell^\pm\ell^\pm\) process}
The \WWll cross-section depends on the $WW$ centre of mass energy $\Mww$.
This variable corresponds to $x_1 x_2 s$ in the EWA approximation formula~(\ref{eq:eff_W_approx}).
In the following we  set  $\Mww \simeq \SI{500}{GeV}$ which
represents a typical value for the invariant mass of the $WW$ system in $pp$ scattering at the $\SI{13}{TeV}$ LHC  (c.f.\ \cite{Fuks:2020att}).

Averaging over $W$ polarisations contracted with the amplitude in \cref{eq:WW Flavour Amplitude} we can find the differential cross section of the $W^\pm W^\pm \to \ell^\pm \ell^\pm$ process which is shown for different mass ratios (viridis colour scheme) in \cref{Fig.
  2HNL diffXSect Massless}.
As expected, the differential cross section decreases as $r_\Nu \to 1$ approximately as
$  {\dd{\sigma}}/{\dd{t}} \propto {(r_\Nu-1)^2}$.
The process fully vanishes in the case of exact HNL pair degeneracy (quasi-Dirac limit).
For $r_\Nu \gg 1$, the differential cross section in the two HNL
model approaches that of a single isolated HNL $\Nu_1$ (dashed orange), so that from
$r_\Nu=10$ the two become indistinguishable at the scales shown here.  This
is due to the fact that for large HNL masses
($m_{\Nu_2}^2\gtrsim |t|_{\max}$) the denominator of the heavier HNL
propagator is always dominated by the mass term, suppressing all $\Nu_2$
contributions by at least $r_\Nu^{-1}$.
For all $r_\Nu$, the numerical value of the cross section is of the same order of magnitude for all $t$.
This corresponds to an evenly distributed angular dependence of the ejectiles.
However, we do observe that for a \qdl pair closer to degeneracy, there is a slight tendency toward back-to-back scattering.
The opposite is true for a highly hierarchical configuration.

\begin{figure}[!t]
  \centering
  \includegraphics[scale=1,valign=c]{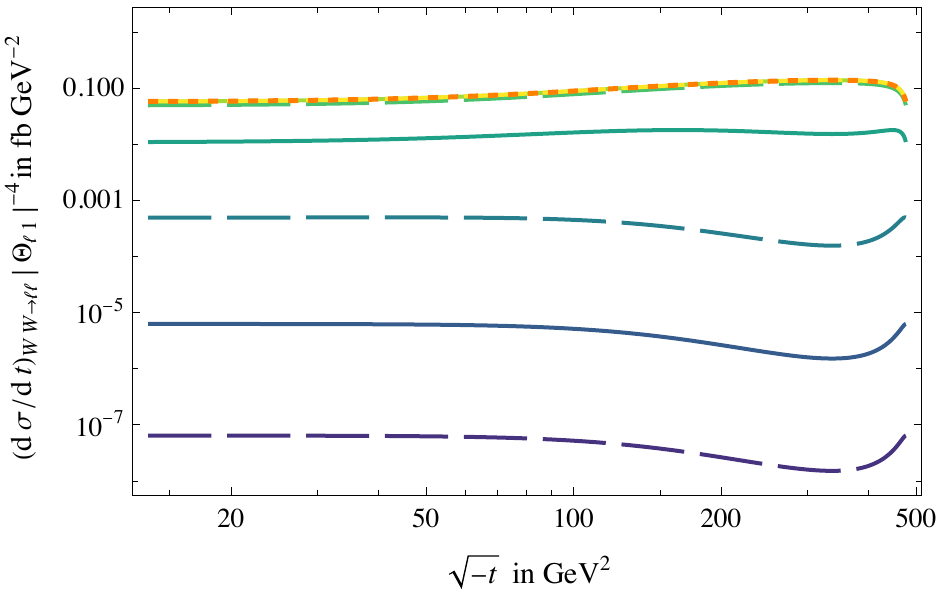}
  \includegraphics[scale=0.92,valign=c]{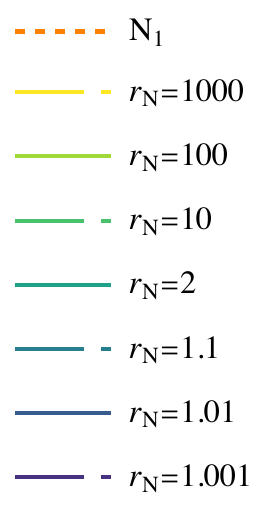}
  \caption{Differential cross section of \WWll scattering in a model with two HNLs with cancelling contributions to the light neutrino masses (\qdl).
    The lighter HNL mass is kept at $m_{\Nu_1}=\SI{150}{GeV}$, while the heavier one is determined by the mass ratio $r_\Nu$, see \protect\cref{eq:rN}.
    The  mixing angle suppression $|\mix_{\ell1}|^4$ is factored out of the result.
    The mass of the final leptons $\ell$ is neglected.
    The centre of mass energy is \SI{500}{GeV}.}
  \label{Fig. 2HNL diffXSect Massless}
\end{figure}

\begin{figure}[!t]
  \centering
  \includegraphics[scale=1,valign=c]{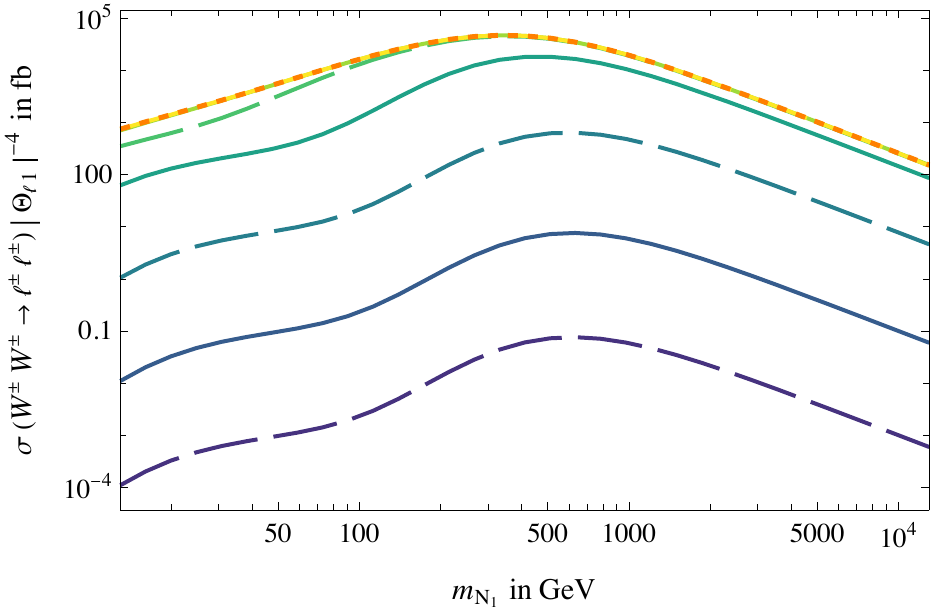}
  \includegraphics[scale=0.92,valign=c]{Figures/Plots/Legends/2HNL_Degeneracy1N.pdf}
  \caption{Total cross section of \WWll scattering in a model with two HNLs
    with cancelling contributions to the light neutrino masses (\qdl).
    The cross section is divided by $|\mix_{\ell 1}|^4$.
    The mass of the lighter HNL is shown as the x-axis, while the heavier one is $r_\Nu$ times larger.
    The masses of the final leptons $\ell$ are neglected.}
  \label{fig:2HNL_xSect_Massless}
\end{figure}

\Cref{fig:2HNL_xSect_Massless} shows the total cross section of \WWll as a function of the lighter HNL mass $m_{\Nu_1}$.
For masses $m_{\Nu_1}\lesssim\Mww/10\simeq\SI{50}{GeV}$ we see a difference between hierarchical HNLs with $r_\Nu \sim 10$ and the single HNL case.
This is due to $m_{\Nu_2}$ being smaller than $\sqrt{s_\ww}$ in this regime, so that $m_{\Nu_2}^2\lesssim |t|_{\max}$ and decoupling of a heavier HNL does not happen.
Hence, a cancellation between the two contributions can occur.

The overall shape of the total cross section, as a function of mass $m_{\Nu_1}$, exhibits maximal value for  $m_{\Nu_1} \sim \Mww$.
Trends for small and large $m_{\Nu_1}$ can be directly understood from the amplitude in \cref{eq:WW Flavour Amplitude} of the \WWll process, which at small mass scales as $m_{\Nu_1}$.
At large masses it is suppressed as $m_{\Nu_1}^{-1}$, which can also be observed
in the cross section.

We note that the cross section in a realistic \qdl model can be substantially larger than that in a single HNL model with the mixing angles close to the seesaw line  (see~\Cref{appendix:1-hnl-case}).

\section{Results: ``neutrinoless double-beta decay''  at Colliders}
\label{sec:pp_lljj}

Using the effective $W$ approximation, the results of the previous section are translated into $pp$ level cross sections.
Although our results were obtained in the \qdl model (i.e.\ in the limit where HNLs' contribution to neutrino masses cancels exactly even if $m_{\Nu_1} \neq m_{\Nu_2}$), they are valid in realistic 2HNL models with non-zero neutrino masses.
In \cref{appendix:CasasIbarra}, it is demonstrated that far from the seesaw line ($|\mix_{\ell I}| \gg |\mix|_{\text{seesaw}}$, the light neutrino's contributions to the process become negligible compared to those of the HNLs.
As we will see, only HNLs with large mixing angles could lead to detection of WBF at the LHC or FCC-hh, so that it is justified to consider the \qdl model.\footnote{We also repeated our analysis for a single HNL in a model ignorant of light neutrino masses.
  At $WW$ level, our results match those of \protect\cite{Fuks:2020att} while at $pp$ level they differ by a factor of  $\sim 1.6$ in the region of full validity according to \cref{appx:effectiveWApprox}.
  The origin of this factor could be related to known limitations of EWAs, see \protect\cite{Ruiz:2021tdt} for detailed discussion.}
This makes numerical integration of the relevant matrix elements simpler and numerically stable.

\subsection{Large Hadron Collider}
\label{sec:LHC}

\begin{figure}[!t]
  \centering
  \includegraphics[valign=c]{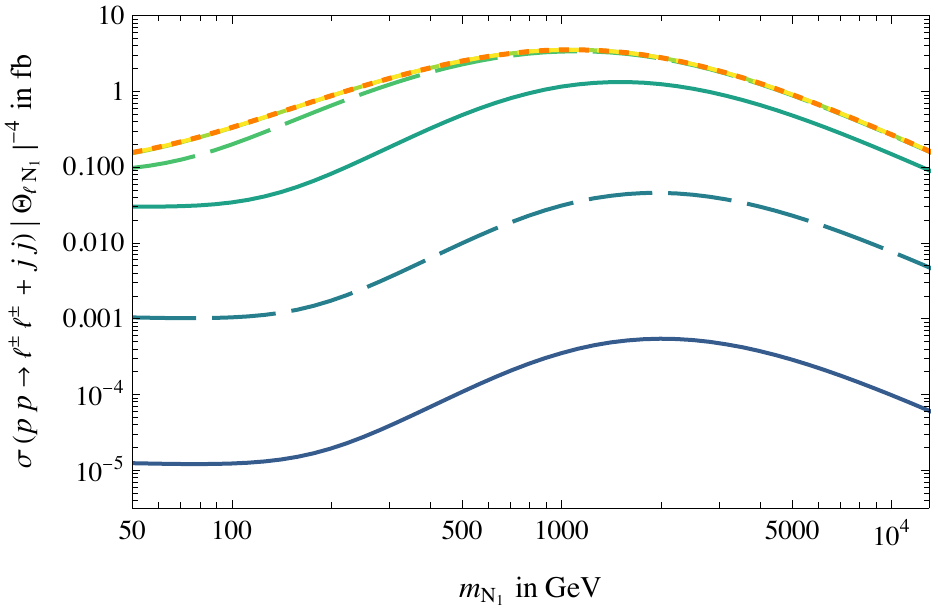}
  \includegraphics[scale=0.92,valign=c]{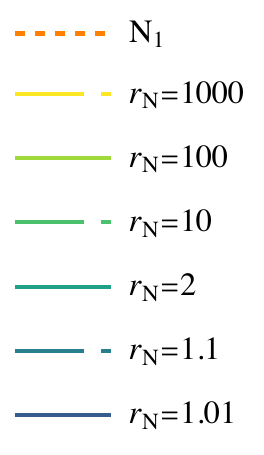}
  \caption{Cross section of \ppll as calculated in the effective $W$ approximation in the model with two HNLs whose mass ratio is equal to $r_\Nu$.
    The mass of the lighter HNL is shown as the x-axis, while another one is $r_\Nu$ times heavier.
    Notations are the same as in \protect\cref{fig:2HNL_xSect_Massless}.
    The dependence on the mixing angle $|\mix_{\ell1}|^4$ is factored out.
    Contribution of the light neutrinos is neglected (see \protect\cref{sec:WW_ll_2HNLs} for details).
    The centre of mass energy is assumed to be $\sqrt{s_\lhc}=\SI{13}{TeV}$.}
  \label{Fig. pp 2HNL xSect Massless}
\end{figure}

\Cref{Fig. pp 2HNL xSect Massless} shows the expected total cross section of $pp\to\ell^+\ell^+ + jj$ for a centre of mass energy $\sqrt{s_\lhc}=\SI{13}{TeV}$.
We remind that the effective W approximation largely underestimates the cross-section for masses \emph{below few hundred GeV} (see \cref{appx:effectiveWApprox} below) and therefore our results only apply for larger masses.
Compared to \cref{fig:2HNL_xSect_Massless} we see that the polarisation decomposition and subsequent folding with the $W$ PDFs  shifts the maximum from around $m_{\Nu_1}\sim\SI{0.5}{TeV}$ to $m_{\Nu_1}\sim\SI{1}{TeV}$.
More importantly, the $pp$ cross section is about 4 orders of magnitude smaller than that of the $WW$.
Beyond this, the general characteristics of the $WW$-scattering case are directly translated into the $pp$ case.
Using the $pp$ cross sections from \cref{Fig.
  pp 2HNL xSect Massless}, we can estimate how many LNV events could be expected at the LHC.
To this end, we multiply the cross section by the maximal \emph{admissible value} of the mixing angles (see \cref{appx:type-I-seesaw} for details).
Namely, \emph{each mixing angle} $\mix_{\ell 1}$ and $\mix_{\ell 2}$ should obey the condition
\begin{equation}
  \bigl|\mix_{\ell I}\bigr|\le\left\{
    \begin{aligned}
      1 &,&m_{\Nu_I}<\mathrm{vev}\\
      \frac{\mathrm{vev}}{m_{\Nu_I}} &,&m_{\Nu_I}>\mathrm{vev}
    \end{aligned}\right.
\label{eq:Theta_max}
\end{equation}
for its respective mass.
Notice that for $m_{\Nu_2} > \mathrm{vev}$ and $r_\Nu > 1$, the perturbativity condition of the HNL $\Nu_2$ provides the most stringent theoretical upper bound on $|\mix_{\ell 1}|$ (under the assumption of \crefrange{eq:toy_2HNL_model}{Eq. Masslessness Condition}).\footnote{This, in particular, demonstrates that decoupling of one of the HNLs, while keeping neutrino masses small is not possible -- the condition for the theory to remain perturbative imposes constraints on the mixings of non-decoupled lighter HNLs.
This is an example of the known violation of the ``decoupling theorem'' \cite{Appelquist:1974tg} in \typeI model, see e.g.\ recent discussion in \cite{Calderon:2022alb}.}

\begin{figure}[!t]
  \centering
  \includegraphics[valign=c]{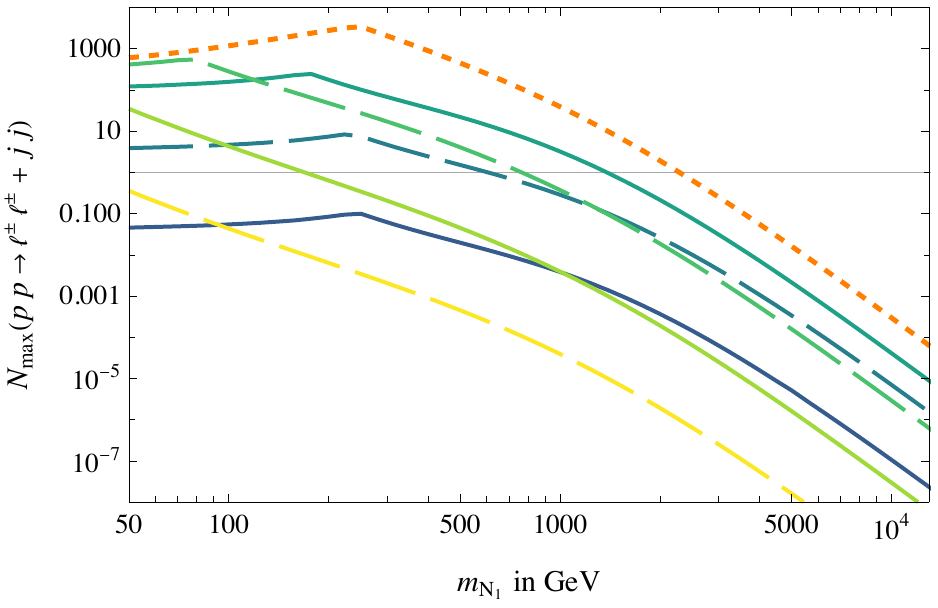}
  \includegraphics[scale=.92,valign=c]{Figures/Plots/Legends/2HNL_Degeneracy1NMod.pdf}
  \caption{The maximal number of WBF events as calculated in the effective $W$ approximation in the model with two HNLs with mass ratio $r_\Nu$.
    The mixing angle $|\mix_{\ell I}|^4$ is maximised, according to \protect\cref{eq:Theta_max}.
    Notice that for $m_{\Nu_2} = r_\Nu m_{\Nu_1} > \mathrm{vev}$, it is the condition on the mixing angle $|\mix_{\ell 2}|$ that dominates.
    The overall efficiency of the signal is assumed $100\%$.
    The centre of mass energy is $\sqrt{s_\lhc}=\SI{13}{TeV}$ and the luminosity $\SI{3}{\per\atto\barn}$.}
  \label{Fig. pp 2HNL NMax Massless}
\end{figure}

\Cref{Fig. pp 2HNL NMax Massless} shows the expected maximal amount of events $N_{\max}$ at the HL-LHC with an integrated luminosity of $\SI{3000}{\per\femto\barn} \equiv \SI{3}{\per\atto\barn}$~\cite{HLLHCDesignReport2020} as a function of the lighter of two HNL masses $m_{\Nu_1}$ for different mass ratios $r_\Nu$, as well as for a single HNL case (dashed orange).
The number of events reaches its maximum at $m_{\Nu_1} \sim \mathrm{vev}/r_\Nu^{1/2}$.
For models with $m_{\Nu_1}<\SI{100}{GeV}$, $N_{\max}$ is the largest for $r_\Nu \gtrsim \mathcal{O}\left(10\right)$, while for $m_{\Nu_1}>\SI{100}{GeV}$ $N_{\max}$ is largest for $r_\Nu \sim \few$.
Notice that the differential cross section for $r_N \ge 2$ looks similar to that of the single HNL.
Therefore, efficiencies of cuts will be similar to those estimated in
\cite{Fuks:2020att}.\footnote{For masses $m_{\Nu_1}\sim\few\,\si{TeV}$ this corresponds to a signal loss of around $\SI{40}{\%}$. }
We stress that for all mass ratios $N_{\max}<1$ for $m_{\Nu_1}\ge
\SI{1.2}{TeV}$ and that the region above $\SI{1}{TeV}$ can only be
probed for $2\le r_\Nu\lesssim 10$.

Furthermore, we can estimate $95\%$~CL exclusion in the $\left(\mix_{\ell\Nu},r_\Nu,m_\Nu\right)$ parameter space.
Below we report our results for $r_\Nu=2$ and $r_\Nu=10$, as they represent the most promising mass ratio regime according to \cref{Fig.
  pp 2HNL NMax Massless}.
Bounds for $r_\Nu<2$ quickly deteriorate $|\mix_{\ell\Nu}|^2_{95\%} \propto |r_\Nu-1|^{-1}$.
For $r_\Nu>10$ growing fraction of the parameter space becomes
excluded due to perturbativity constraints,~\Cref{sec:perturbativity-limit}.
In a model with $r_\Nu\sim\few$ (\cref{Fig.
  pp 2HNL U2 Massless 2}), and $100\%$ efficiency
($\epsilon\sim 1$, yellow), we find that a WBF search at the HL-LHC can probe HNL masses $m_{\Nu_1}$ up to $\SI{1}{TeV}$.
This mass range is limited to $m_{\Nu_1}\lesssim\SI{500}{GeV}$ for a lower efficiency ($\epsilon\sim 0.1$, dashed grey).
In a model with $r_\Nu\sim\mathcal{O}\left(10\right)$ (\cref{Fig.
  pp 2HNL U2 Massless 10}) the cross section per $|\mix_{\ell 1}|^4$ is higher, but the range of accessible masses is smaller, due to the perturbativity condition of \cref{eq:Theta_max}.
As a result the maximally probed HNL mass is $m_{\Nu_1}\lesssim\SI{500}{GeV}$.
This limit drops to around $\SI{300}{GeV}$ for a lower efficiency.
\begin{figure}[!t]
  \centering
  \subfloat[][$r_\Nu=2$\label{Fig. pp 2HNL U2 Massless 2}]{\includegraphics[scale=.791,valign=c]{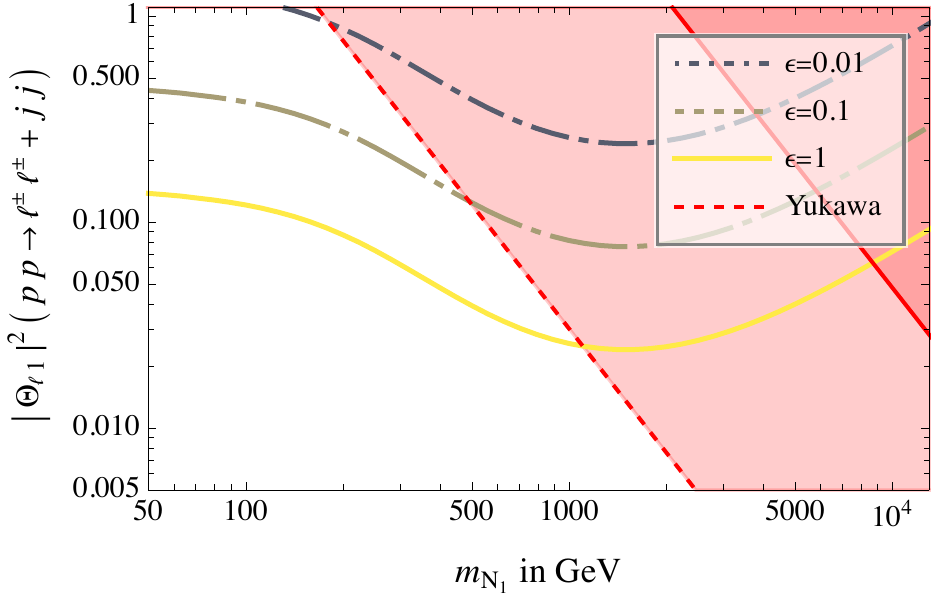}}
  \subfloat[][$r_\Nu=10$\label{Fig. pp 2HNL U2 Massless 10}]{\includegraphics[scale=.791,valign=c]{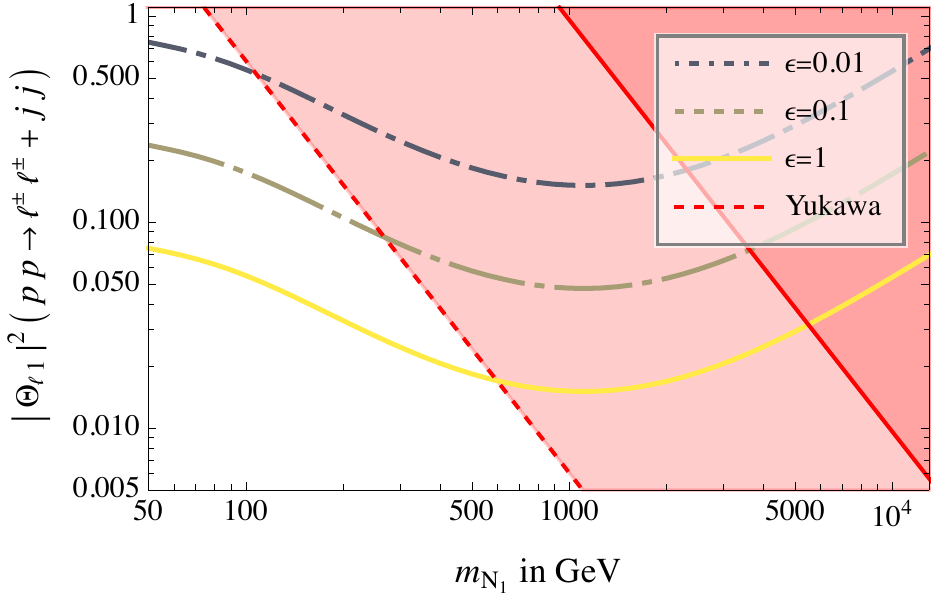}}
  \caption{\emph{Optimistic} exclusion limits on the $\Nu_1$ parameters via WBF-mediated \ppll events in the model with two HNLs with mass ratio $r_\Nu$.
    The dashed red line indicates the perturbativity limit of the Yukawa coupling $|F_{\ell2}|=1$, while the solid red line indicates $|F_{\ell2}|=4\pi$.
    The exclusions limits are $\SI{95}{\%}$ CL assuming zero background.
    The centre of mass energy is $\sqrt{s_\lhc}=\SI{13}{TeV}$ and the luminosity $\SI{3}{\per\atto\barn}$.}
  \label{Fig. pp 2HNL U2 Massless}
\end{figure}
With efficiency dropping to $\epsilon\sim\mathcal{O}\left(10^{-2}\right)$, a signal cannot be expected for any meaningful $\left(\mix_{\ell\Nu},r_\Nu,m_\Nu\right)$ combination.

\subsection{Future Circular Collider}
\label{sec:FCC}

Sensitivity of the WBF process is maximal for $\sqrt{s_\ww} \sim m_{\Nu_1}$.
The former increases with the $\sqrt s$ of $pp$ collision.
Therefore the process under consideration can be expected to be more efficient at the Future Circular Collider in the pp mode (FCC-hh) \cite{FCC:2018byv}.
Using its projected centre of mass energy $\sqrt{s_\fcc}=\SI{100}{TeV}$ and machinery, developed in this work, we can estimate the cross section of the \ppll process, shown in \cref{Fig.
  pp 2HNL xSectFCC Massless}.

\begin{figure}[!t]
  \centering
  \includegraphics[valign=c]{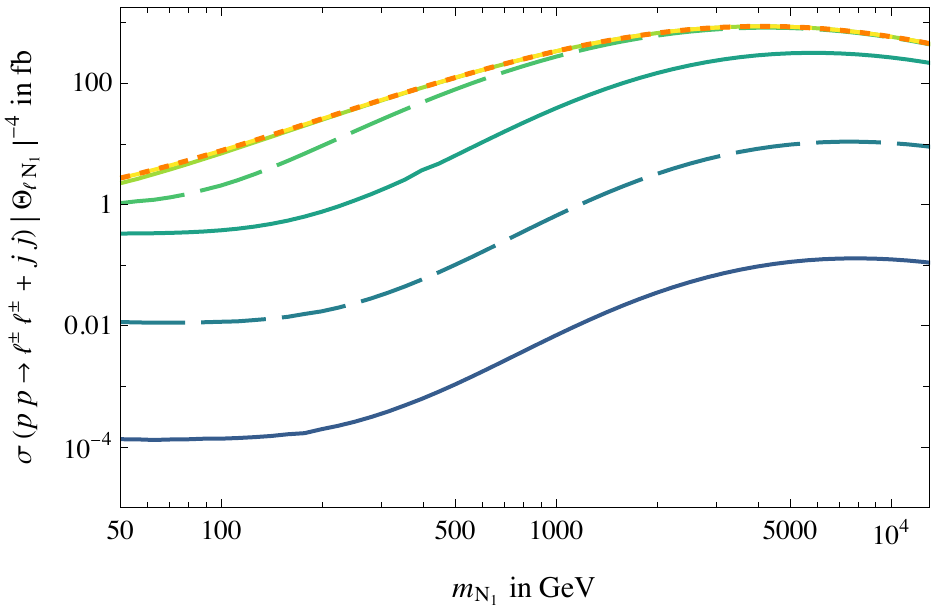}
  \includegraphics[scale=0.92,valign=c]{Figures/Plots/Legends/2HNL_Degeneracy1NMod.pdf}
  \caption{Cross section of \ppll as calculated in the effective $W$ approximation in the model with two HNLs whose mass ratio is equal to $r_\Nu$.
    The mass of the lighter HNL is shown as the x-axis, while another one is $r_\Nu$ times heavier.
    Notations are the same as in \protect\cref{fig:2HNL_xSect_Massless}.
    The dependence on the mixing angle $|\mix_{\ell 1}|^4$ is factored out.
    Contribution of the light neutrinos is neglected (see \protect\cref{sec:WW_ll_2HNLs} for details).
    The centre of mass energy is assumed to be $\sqrt{s_\fcc}=\SI{100}{TeV}$.}
  \label{Fig. pp 2HNL xSectFCC Massless}
\end{figure}

We find that the cross section reaches its maximum at $m_{\Nu_1}\simeq\SIrange{2}{3}{TeV}$ (compared to $m_{\Nu_1}\simeq\SIrange{1}{1.5}{TeV}$ at LHC energies).
It is, furthermore,  $\sim2.5$ orders of magnitude larger at FCC energies and reaches $\sim\SI{1}{\pico\barn}$ for a hierarchical HNL pair.

\begin{figure}[!t]
  \centering
  \includegraphics[valign=c]{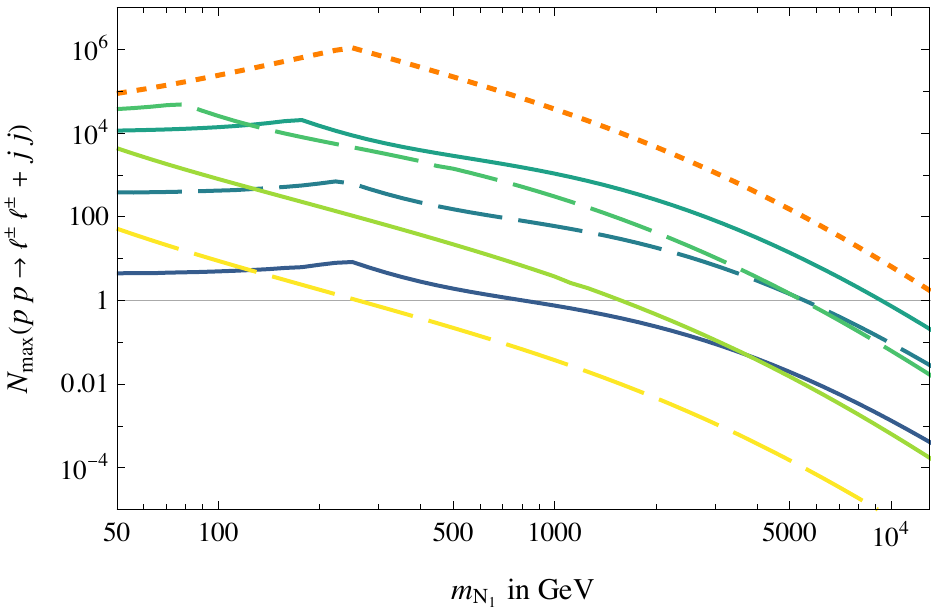}
  \includegraphics[scale=.92,valign=c]{Figures/Plots/Legends/2HNL_Degeneracy1NMod.pdf}
  \caption{The maximal number of WBF events as calculated in the effective $W$ approximation in the model with two HNLs with mass ratio $r_\Nu$.
    The mixing angle $|\mix_{\ell 1}|^4$ is maximised, according to \protect\cref{eq:Theta_max}.
    Efficiency of the detector is assumed $100\%$.
    The centre of mass energy is $\sqrt{s_\fcc}=\SI{100}{TeV}$ and the luminosity $\SI{30}{\per\atto\barn}$.}
  \label{Fig. pp 2HNL NMaxFCC Massless}
\end{figure}

With a target integrated luminosity $\SI{30}{\per\atto\barn}$\cite{Aleksa:2019pvl}, this corresponds to a maximal theoretically admissible event number $N_{\max}$ as shown in \cref{Fig. pp 2HNL NMaxFCC Massless}.
Again, we see that for $m_{\Nu_1}>r_\Nu^{-1/2}\mathrm{vev}$ Yukawa perturbativity becomes the main theoretical constraint on the mixing angle $\mix_{\ell 1}$, resulting in $r_\Nu\sim\few$ to yield the largest $N_{\max}$.

\begin{figure}[!t]
  \centering
  \subfloat[][$r_\Nu=2$\label{Fig. pp 2HNL U2FCC Massless 2}]{\includegraphics[scale=.791,valign=c]{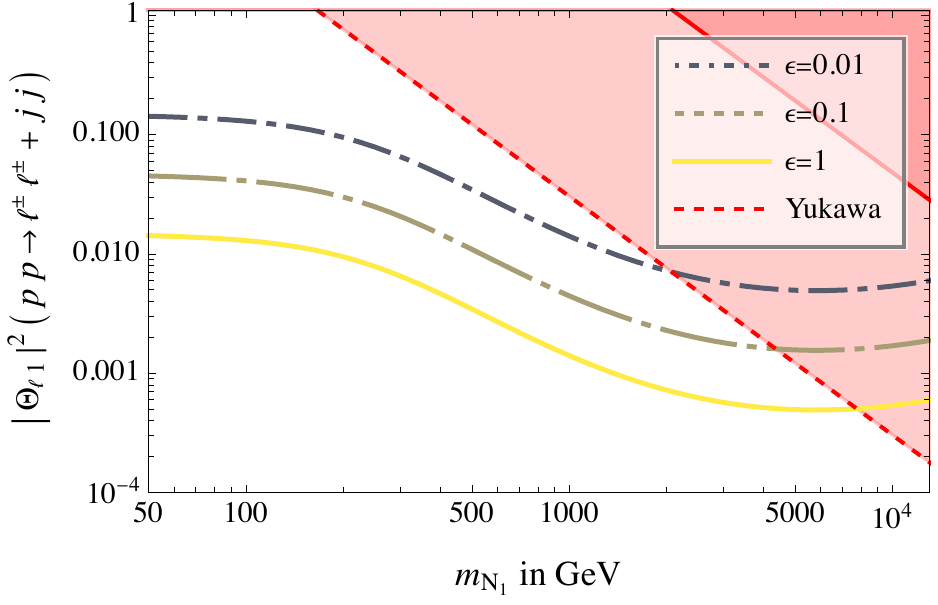}}
  \subfloat[][$r_\Nu=10$\label{Fig. pp 2HNL U2FCC Massless 10}]{\includegraphics[scale=.791,valign=c]{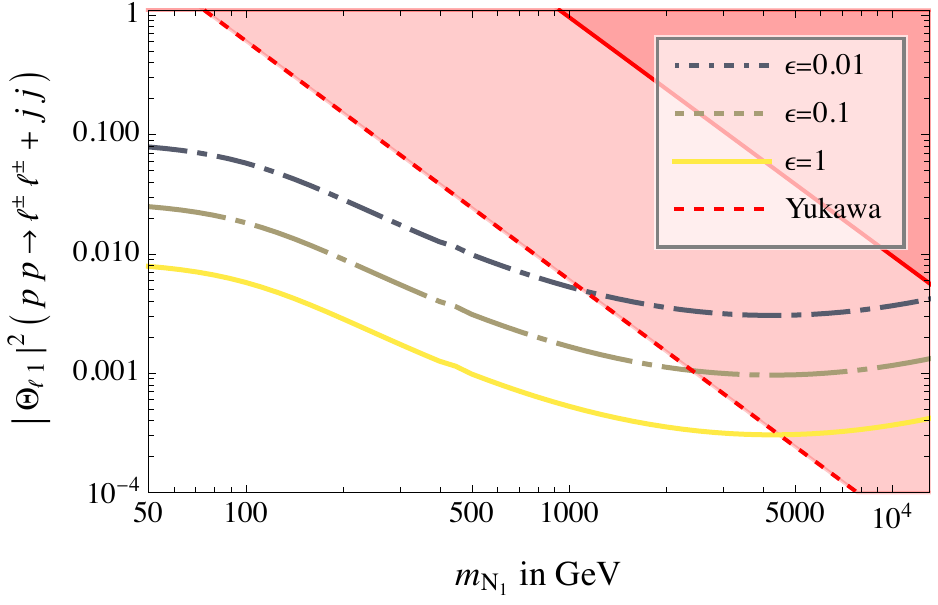}}
  \caption{\emph{Optimistic} exclusion limits on the $\Nu_1$ parameters via WBF-mediated \ppll events in the model with two HNLs with mass ratio $r_\Nu$.
  The dashed red line indicates the perturbativity limit of the Yukawa coupling $|F_{\ell2}|=1$, while the solid red line indicates $|F_{\ell2}|=4\pi$.
  The limits are obtained for \SI{95}{\%} CL exclusions assuming zero background.
  The assumed detector efficiency is $\epsilon$, the centre of mass energy $\sqrt{s_\fcc}=\SI{100}{TeV}$ and the luminosity $\SI{30}{\per\atto\barn}$.}
  \label{Fig. pp 2HNL U2FCC Massless}
\end{figure}

Due to the significantly larger cross section and higher integrated luminosity, the relevant  $\left(\mix_{\ell\Nu},r_\Nu,m_\Nu\right)$ parameter space covered by \ppll at the FCC-hh opens up.
For both $r_\Nu=2$ and $r_\Nu=10$ (see \cref{Fig. pp 2HNL U2FCC Massless 2,,Fig. pp 2HNL U2FCC Massless 10} respectively), physically relevant mixing angles $\mix_{\ell\Nu}$ are within reach for a \qdl HNL pair with $m_{\Nu_1}\sim\few\,\si{TeV}$ assuming a detection efficiency $\epsilon \ge 0.1$.

\section{Discussion and conclusion}
\label{sec:discussion}

In this work we analysed the collider probe for Majorana particles with masses ranging from $\sim\SI{50}{GeV}$ to $\sim \SI{5}{TeV}$.
The process that we considered is a direct analog of the \nubb decay --- \ppll (\cref{fig:main_process}).
Compared to previous works on the subject (c.f.\ \cite{Dicus:1991fk,Ali:2001pk,Panella:2001wq,Ng:2015hba,Fuks:2020att})
we concentrated on models where HNLs are solely responsible for generating neutrino masses.
All Majorana particles --- HNLs and active neutrinos --- contribute to the process in question and we analysed their interference.
We demonstrated that such an interference is necessarily \emph{destructive} as a consequence of smallness of neutrino masses, as compared to other relevant energy scales.
There are two limiting regimes depending on whether HNLs' mixing angles $|\mix_{\ell \Nu}|$ are comparable or much larger than the naive ``seesaw limit'' $|\mix|_{\text{seesaw}}^2 = \frac{\sqrt{|\Delta m_{\text{\rm atm}}^2|}}{m_{\Nu}}$ (see text around \cref{eq:1})
\begin{enumerate}
\item For $|\mix_{\ell\Nu}|\sim |\mix|_{\text{seesaw}}$ cancellation between HNLs and neutrino states can occur and the resulting cross section is proportional to $m_{\nu_i}^2$.
\item For $|\mix_{\ell\Nu}|\gg |\mix|_{\text{seesaw}}$ (while still keeping neutrino masses small, as experimentally observed) the situation is quite different.
  The contribution of active neutrinos is negligible, but cancellation between HNL states does occur (similarly to the way it happens in the neutrino mass matrix, see e.g.\ \cite{Kersten:2007vk}).
  On the one hand, the cross-section is always smaller than that of a single HNL with the same mixing.
  On the other hand, the cross-section can get enhanced as compared to the naive scaling of $\sigma\propto m_\nu^2$ than could have occurred from the Weinberg operator per se (c.f.\ \cite{Fuks:2020zbm}).
  This enhancement is roughly by a factor of $|\mix_{\ell\Nu}|^4/ |\mix|_{\text{seesaw}}^4$ and occurs only if HNLs are sufficiently far from the quasi-Dirac limit (i.e.\ $m_{\Nu_2} /m_{\Nu_1} = r_\Nu \gtrsim \text{few}$).
\item Lastly, large mass splitting $r_\Nu \gg 1$ does not allow to recover the limit of a single HNL (as considered, e.g.\ in \cite{Fuks:2020att}).
  Naively, in this case the heavier HNL $\Nu_2$ and its contribution to \ppll should disappear.
  This is not the case in our realistic (and UV-complete) model (\typeI with 2 HNLs).
  Indeed, perturbativity demands that all Yukawa couplings are smaller than~$\sim 1$.
  For HNLs heavier than the Higgs's VEV this condition together with the requirement of the smallness of neutrino masses caps not only the value of the mixing angle $|\mix_{\ell 2}|$ but by extension also $|\mix_{\ell 1}|$.
\end{enumerate}
As a result, the \ppll searches are most sensitive for $r_\Nu \sim \text{few}$ at $m_{\Nu_1} \sim \SIrange{0.5}{1}{TeV}$.\footnote{Contribution of the hierarchical HNLs to the \nubb process have been studied recently in \protect\cite{Asaka:2021hkg,Asaka:2020lsx,Asaka:2020wfo}.}
Even in this case the cross section for the process \ppll is about 1 order of magnitude smaller than that of a single HNL with the same mixing angle.
At the LHC, the best exclusion limit obtainable in this case would reach $|\mix_{\ell 1}|^2 \sim \SIrange{0.02}{0.05}{}$ (assuming background-free search, $100\%$ efficiency of the signal detection and the luminosity of $\SI{3}{\per\atto\barn})$.
In particular, even at the end of the high-luminosity LHC phase (\SI{3}{\per\atto\barn} integrated luminosity) these limits are non-competitive with those coming from the \emph{electroweak precision tests} (EWPT) \cite{Fernandez-Martinez:2016lgt} or from non-observations of lepton \emph{flavour violating} process, involving charged leptons \cite{Calderon:2022alb}.
At the FCC-hh, the exclusion limit could be as low as $|\mix_{\ell 1}|^2 \sim (\SIrange{2}{5}{})\times 10^{-4}$ for $r_\Nu\sim\few$ in a mass range of $m_{\Nu_1} \sim \few\,\si{TeV}$.
These bounds are competitive with EWPT exclusion.
Furthermore, in the regime of $m_{\Nu_1}\sim\few\,\si{TeV}$, they
could improve on exclusion bounds set by those of
$\mathrm{CCDY}+W\gamma$ searches \cite{Pascoli:2018heg}.

Throughout this paper we assumed background free searches.
This is true only approximately.
Indeed, there are Standard Model processes that may lead, e.g.\ to the appearance of two same-sign $W$'s in the final states, with subsequent $W\to \ell\nu$ decays with low missing transverse momentum, see e.g.\ \cite{Alboteanu:2008my,Frandsen:2009fs}.
Additional sources of the background include: mis-interpretation of jets as leptons (so-called \emph{fake leptons}) \cite{ATLAS:2010vza,ATLAS:2014ffa,Thusini:2017ztk}; processes with 3 or more leptons, some of them escaping the detection; processes with the charge misidentification, etc.
At the same time, previous searches \cite{CMS:2012ggh,CMS:2022rqc} show that the backgrounds can indeed be made low by choosing suitable discriminating variables.
We leave the detailed studies to the future work.
However, we do not expect a drastic change of the results, as further reduction of the signal due to cuts  \cite{Fuks:2020att} will be in the same ballpark as underestimation of the signal due to the EWA.

\textbf{In summary:} We explored the potential of the WBF signal at the LHC.
We concluded that for the planed HL-LHC upgrade it can only contribute meaningful mixing bounds in a quasi-Dirac-like HNL model with HNL mass $m_{\Nu_1}<\SI{1}{TeV}$ and the corresponding mass $m_{\Nu_2}$ according to the mass ratio.
For the FCC-hh, the mass range can be extended to $m_{\Nu_1}\sim\few\,\si{TeV}$ yielding competitive limits for mass ratios of $r_\Nu\sim\few$ (see \cref{fig:ExclusionLimitResults}).

The already existing bounds derived from the WBF process \cite{CMS:2022rqc} can be reinterpreted as bounds on a two HNL model in the highly hierarchical limit
owing to the fact that differential cross section for $r_\Nu \gg 1$ looks similar to the single HNL case and therefore the signal acceptance stays roughly the same.
We stress again that this does not correspond to a decoupling in the traditional sense, as the ``decoupled'' and ``non-decoupled'' HNLs' mixing angles are explicitly related by the requirement that neutrino masses remain small as observed experimentally.

\subsection{Relation between LNV effects and small  neutrino masses}
\label{sec:equiv-betw-lnv}

Let us comment on the suppression of LNV effects as a consequence of smallness of neutrino masses, a subject of discussion in many works in the past \cite{Shaposhnikov:2006nn,deGouvea:2007qla,Kersten:2007vk,Moffat:2017feq,Tastet:2019nqj,Drewes:2019byd}.
The relation between the two signals would, essentially, mean that every LNV effective operator \cite{deGouvea:2007qla} is dependent on the Weinberg operator~\cite{Weinberg:1979sa}.
Our results demonstrate that this is not the case, as even in the \qdl model, where neutrino mass is exactly zero, the WBF signal remains finite.
We do see, however, that the destructive nature of HNL interference is related to the cancellation of their contributions to the neutrino masses.
In the Dirac-limit ($r_\Nu \to 1$ in our terms), both the smallness of neutrino masses and LNV collider effects would be proportional to the same small perturbation (e.g.\ $\Delta M/M$, \cite{Shaposhnikov:2006nn, Kersten:2007vk}).
However, in the hierarchical case ($r_\Nu > \few$), this does not apply and the two kinds of LNV operators become independent.

The condition $r_\Nu > \few$ implies that our results are not directly applicable to models like the $\nu$MSM (where HNLs are highly degenerate~\cite{Asaka:2005pn,Asaka:2005an,Shaposhnikov:2006nn}, see \cite{Boyarsky:2009ix} for review).
Rather, these results are applicable for leptogenesis models with 3 HNLs~\cite{Akhmedov:1998qx,Drewes:2017zyw,Abada:2018oly} where the ratio $r_\Nu \sim \few$ and large mixing angles are consistent with successful leptogenesis.
We leave the generalisation of our results for the case of 3 HNLs to future works. 

The hierarchical HNL spectrum has its own drawbacks.
As pointed out in past literature \cite{Pilaftsis:1991ug,Korner:1992zk,Kersten:2007vk,AristizabalSierra:2011mn,Lopez-Pavon:2012yda,Pascoli:2013fiz,Lopez-Pavon:2015cga, Haba:2016lxc,Drewes:2019byd}, the light neutrino mass states generated in models with a highly hierarchical \qdl HNL pair can exhibit a significant running of the light masses.
This makes these kinds of models theoretically less appealing.
Owing to the fact that even hierarchical HNLs are not expected to provide a sizeable signal at the HL-LHC, we leave the detailed exploration of this question to the future.

\acknowledgments
We are grateful to Richard Ruiz and Benjamin Fuks for providing fruitful feedback on the initial version of the manuscript and to Mads T.\ Frandsen for his valuable input on the JLS.'s Master's thesis~\cite{Jonathan_Schubert_MSc}.
We would also like to thank Kevin Urqu\'ia Calder\'on, Inar Timiryasov, as well as members of the ``NBI HNL group''
for many discussions during this work.
This project has received funding from the European Research Council (ERC) under the European Union's Horizon 2020 research and innovation programme (GA 694896) and from the Carlsberg Foundation.
O.R.\ would like to thank the Instituto de Fisica Teorica (IFT UAM-CSIC) in Madrid for support via the Centro de Excelencia Severo Ochoa Program under Grant CEX2020- 001007-S, during the Extended Workshop ``Neutrino Theories'', where this work developed.

\bibliographystyle{JHEP}%
\bibliography{PaperLiterature}

\begin{appendices}
  \crefalias{section}{appsec}

  \section{Type I seesaw}\label{appx:type-I-seesaw}

  In this section we summarise the main formulas of the \typeI model with
  the goal to fix the notations. For references see, e.g.
  \cite{Alekhin:2015byh,Abdullahi:2022jlv} and refs.\ therein.

  Before Electro-Weak Symmetry Breaking (EWSB) the \typeI Lagrangian
  is
  \begin{align}\label{Eq:Lagrangian_seesaw_preEWSB}
    \mathcal{L}_\mathrm{seesaw} &= \mathcal{L}_\mathrm{SM} + \frac{i}{2}\nu_{R I}^{\dagger}\bar{\sigma}^{\mu}\partial_{\mu}\nu_{R I} - \left(F_{\alpha I}\right)^*\left(L_{\alpha}\cdot\tilde{\phi}\right) ^{\dagger}\nu_{\R I} - \frac{M_I}{2}{\nu}_{R I}^T\nu_{R I} + h.c.
  \end{align}
  Here $\nu_{RI}$ are the new right-chiral singlet states with $I=1,\dots,\mathcal{N}$ with associated Majorana masses $M_I$, $L_{\alpha}$ is the SM left chiral SU(2) doublet $L_{\alpha}=\begin{pmatrix}\nu_{\alpha}\\l_{\alpha}\end{pmatrix}_L$, where $\alpha=e,\mu,\tau$ and $\tilde{\phi}_a=\varepsilon_{ab}\phi_b^*$, where $\phi$ is the Higgs doublet.
  After EWSB, this can be written as $\phi=\frac{1}{\sqrt{2}}\begin{pmatrix}0\\ v\end{pmatrix}$, where $v$ is the vacuum expectation value (VEV).
  And so $\left( L_{\alpha}\cdot \tilde{\phi}\right)=\frac{v}{\sqrt{2}}\nu_{L\alpha}$ after EWSB.
  The terms
  \begin{align}
    \label{eq:8}
    \mathcal{L}_D &= -\left(F_{\alpha I}^{\nu}\right)^*\frac{v}{\sqrt{2}}\nu_{L\alpha}^{\dagger}\nu_{R I}+ h.c.
  \end{align}
  are equivalent to Dirac mass terms with
  $(m_D)_{\alpha I}=\frac{v}{\sqrt{2}}\left(F_{\alpha I}\right)^*$.
  Diagonalising the mass term
  leads to two types of Majorana states: $\mathcal N$ heavy Majorana states $\Nu_I$ with masses
  \begin{equation}
    \label{eq:3}
    m_{\Nu_I}\simeq M_I
  \end{equation}
  and three light Majorana neutrinos whose masses are given by the \emph{seesaw formula}
  \begin{equation}
    \label{eq:seesaw formula}
    \Bigl( V^\dagger \mathrm{diag}(m_{\nu_1},m_{\nu_2},m_{\nu_3}) V \Bigr)_{\alpha\beta} \simeq \sum_I \mix_{\alpha I} \mix_{\beta I} m_{\Nu_I}.
  \end{equation}
  Here $V_{i\alpha}$ is the PMNS matrix, while
  $\mix_{\alpha I}$ are the active-sterile mixing angles\footnote{To make the
  notation more readable, we will sometimes write $\mix_{\ell\Nu}$ instead of
  $\mix_{\alpha I}$.}
  \begin{equation}
    \label{eq:mixing angle yukawa relation}
    \mix_{\alpha I} \equiv \frac{F_{\alpha I} v}{M_I}.
  \end{equation}
  The Majorana scale is not fixed and can range from $\sim\si{eV}$ to $\SI[parse-numbers=false]{10^{15}}{GeV}$.

  \subsection{Perturbativity limit}
  \label{sec:perturbativity-limit}
  The seesaw relation in \cref{eq:seesaw formula} is derived under the assumption
  \begin{equation}
    \label{eq:theta_max_low_mass}
    \text{Type-I seesaw:}\quad |\mix_{\alpha I}| < 1.
  \end{equation}
  However, for large HNL masses, $m_{\Nu_I} \gtrsim v$, the requirement of perturbativity of the model described by \cref{Eq:Lagrangian_seesaw_preEWSB}, $|F_{\alpha I}| < 1$ becomes more restrictive than \cref{eq:theta_max_low_mass} (see also \cite{Chanowitz:1978mv, Durand:1989zs, Fajfer:1998px, Ipek:2018sai,Calderon:2022alb} for the discussion of perturbativity in the Type I seesaw model).
  Indeed, owing to \cref{eq:mixing angle yukawa relation} we get
  \begin{equation}
    \label{eq:perturbativity}
    \text{Perturbativity:}\quad|\mix_{\alpha I}| < \frac{v}{M_I}.
  \end{equation}
  The conditions in \cref{eq:theta_max_low_mass,,eq:perturbativity} were used in \cref{sec:pp_lljj} when deriving maximal number of events produced at colliders \cref{Fig. pp 2HNL NMax Massless}.\footnote{Notice that in a model as described in \cref{sec:WW_ll_2HNLs}, this requirement will be imposed on the heavier of the two HNLs and entail a limit on $\mix_{\ell 1}$ through \cref{eq:toy_2HNL_model}.}
  Owing to the extra loop factor $\frac{1}{(4\pi)^2}$ the perturbativity limit~(\ref{eq:perturbativity}) can be pushed up by a factor of $4\pi$.

  Finally, it should be noted that a \typeI model produces light neutrino masses of the order $m \sim |\mix|^2 m_{\Nu_1}$.
  This defines a minimal mixing angle, admissible for a given HNL mass, commonly referred to as the \emph{seesaw line}
  \begin{equation}
    \label{eq:1}
    |\mix|_{\text{seesaw}}^2 = \frac{\sqrt{|\Delta m_\mathrm{atm}^2|}}{m_{\Nu}}.
  \end{equation}

  \section{Single HNL Case -- Seesaw Line}\label{appendix:1-hnl-case}

  In a toy model including only 1 HNL, the smallness of the light neutrino mass
  states can only be introduced by a sufficiently small mixing
  $\mix_{\alpha \Nu}$.  This model can not account for neutrino oscillations,
  since it can maximally generate a single potentially degenerate, non-zero,
  light mass state level \cite{Drewes:2019mhg}. For demonstration purposes,
  we will only consider a single non-zero light state with a PMNS-like mixing
  of $1$. We will further generously set the value of the
  light neutrino mass to $\SI{1}{eV}$, which roughly corresponds to the current
  direct upper bound on neutrino masses \cite{KatrinBounds2022}. To this effect,
  we consider a mixing $|\mix_{\ell\Nu}^2|=\SI{1}{eV}/m_\Nu$, which yields a
  constant light mass state for a given $m_\Nu$ according to \cref{eq:6}.

\begin{figure}
    \centering
    \includegraphics[scale=0.92,valign=c]{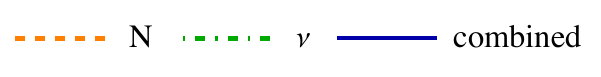}\\
    \vspace{-.2em}
    \includegraphics[scale=0.92,valign=c]{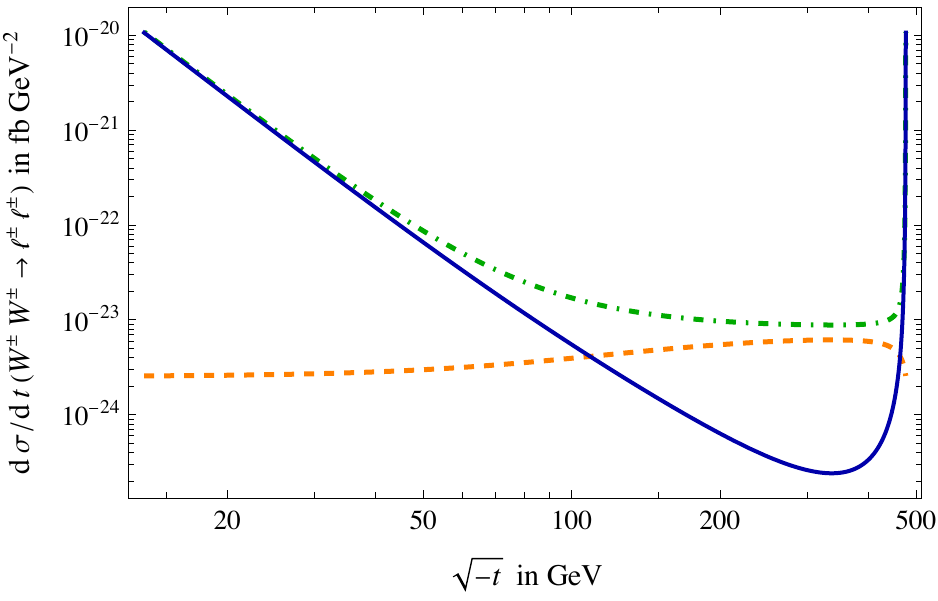}
    \caption{Differential cross section of \WWll scattering in the case of a
      single HNL with mass $m_\Nu=\SI{150}{GeV}$ as a function the
      Mandelstam-variable $t$ at a centre of mass energy of \SI{500}{GeV}.  The
      mixing angle here is a function of the HNL mass, such that
      $|\mix^2_{\ell\Nu}m_\Nu|=\SI{1}{eV}\gtrsim m_{\nu_e}$. For further information see text.}
    \label{Fig. 1HNL diffXSect constLightMass}
  \end{figure}

  Averaging over $W$ polarisations contracted with the amplitude in \cref{eq:WW
    Flavour Amplitude} we find the differential cross section of the
  \WWll process considering both heavy and light
  states.  This is shown for an HNL mass of $\SI{150}{GeV}$ as the blue graph
  in \cref{Fig. 1HNL diffXSect constLightMass}.  Similarly, we can derive the
  differential cross section if only light (shown in dash-dotted green) or
  heavy (dashed orange) mass states contributed to the process.\newline It's
  important to note that the isolated light mass state case yields a larger
  differential cross section than both the isolated heavy case and the combined
  case.  This is true over the entire $t$ range, while at the extremal values
  the isolated $\Nu$-case drops slightly and the combined case approaches the
  $\nu$-case.  That is to say, the isolated heavy case results in a relatively
  even angular distribution, while the light case strongly favours back-to-back
  scattering.  Due to the cancelling nature of the two contributions, which is
  the strongest while both $u$ and $t$ are large when compared to $m_\Nu^2$,
  the combined differential cross section has an even stronger tendency toward
  back-to-back scattering.

  \begin{figure}
    \centering
    \includegraphics[scale=0.92,valign=c]{Figures/Plots/Legends/1HNL_StandardHorizon.pdf}\\
    \vspace{-.2em}
    \includegraphics[scale=0.92,valign=c]{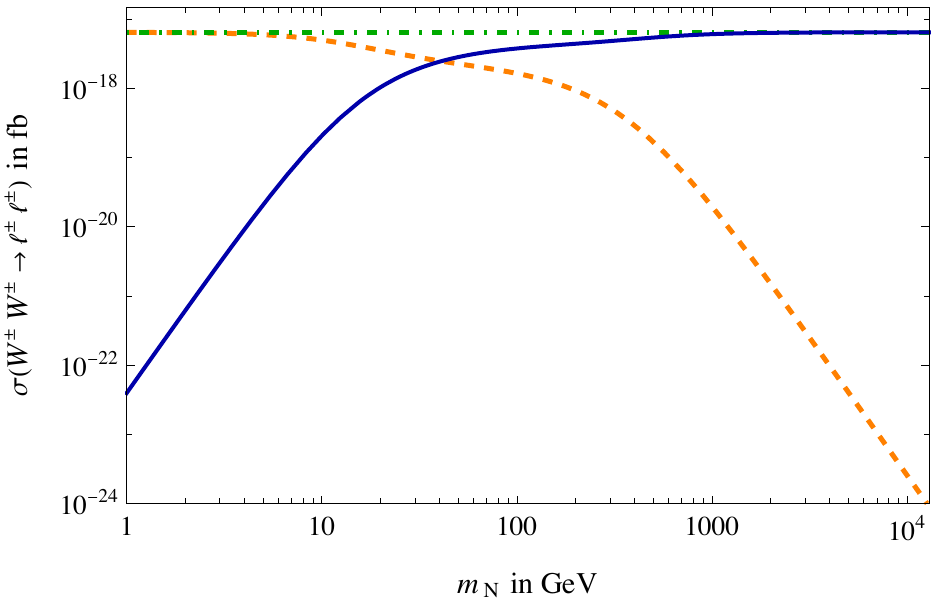}
    \caption{Cross section of \WWll scattering in the case of a single HNL as a
      function of the HNL mass at a centre of mass energy of \SI{500}{GeV}.
      The mixing angle here is a function of the HNL mass as well, such that
      $|\mix^2_{\ell\Nu}m_\Nu|=\SI{1}{eV}\gtrsim m_{\nu_e}$. For further information see text.}
    \label{Fig. 1HNL xSect constLightMass}
  \end{figure}

  Integrating the differential cross section with respect to $t$ we find the
  total cross section, shown in \cref{Fig. 1HNL xSect constLightMass}.  As
  prescribed by the mixing angle, the isolated light mass state case (again in
  dash-dotted green) results in a constant cross section with respect to
  $m_\Nu$.  As expected from the differential cross section the light line
  represents an upper limit to the numerical value of the combined cross
  section (solid blue).  For small HNL masses ($m_\Nu^2\lesssim |t|_{\min}$)
  both heavy and light mass states will have a near identical differential
  cross section, leading to a vanishing combined cross section.  For large HNL
  masses ($m_\Nu^2\gtrsim |t|_{\max}$) the denominator of the HNL propagator is
  always dominated by the mass term.  This, combined with the mixing
  prescription, results in the cross section of the heavy case dropping as
  $m_\Nu^{-4}$, and the combined cross section asymptotically approaching the
  light line.

  We would like to draw attention to the fact that the numerical values of the
  cross section resulting in this model are vanishingly small.  Consequently,
  even given the uniqueness of the signature, there would be no hope of
  detection of a process involving \WWll in terms of a single HNL model.

  \section{Realistic model with two HNLs plus light Majorana
    states}\label{appendix:CasasIbarra}

  Given the importance of the light mass states in the single HNL case, it is
  prudent to investigate their relevance when it comes to the two HNL case as
  well.  An easy way to incorporate them into the model is by using the
  Casas-Ibarra parametrisation of the mixing angles
  \begin{equation}\label{Eq. CasasIbarraMixing}
    \mix = iU\mathrm{diag}(m_{\nu_1},m_{\nu_2},m_{\nu_3})^{1/2}\Omega\mathrm{diag}(m_{\Nu_1},\dots,m_{\Nu_\mathcal{N}})^{-1/2},
  \end{equation}
  which, by design, generates the light mass states $m_{\nu_i}$ through
  \cref{eq:6}.  Here $\Omega$ is an orthogonal $3\times\mathcal{N}$ matrix
  holding all degrees of freedom of the model that aren't fixed by the
  PMNS matrix $U$, and the mass states $m_{\nu_i}$, and $m_{\Nu_I}$.
  To account for neutrino oscillations with 2 HNLs, we require the lightest
  neutrino mass state to be massless.  The most general explicit form of
  $\Omega$ in case of normal hierarchy (NH) is given by
  \begin{equation}\label{Eqn. CasasIbarraParametrisation}
    \Omega =\quad
    \begin{pmatrix}
      0  & 0 \\
      \cos\omega & \sin\omega \\
      -\xi\sin\omega & \xi\cos\omega
    \end{pmatrix},
  \end{equation}
  with $\xi=\pm1$, and $\omega\in\mathbb{C}$.  The parity $\xi$ can be chosen
  by simultaneous redefinition of the fields and $\omega$ so that we will
  choose $\xi=+1$ without loss of generality.  In the following we present the
  numerical values for the case of normal hierarchy with $\ell=e$.\footnote{The
    results will be equivalent for other choices up to orders of few in
    PMNS matrix entry ratios.}  The real part of the Casas-Ibarra
  parameter $\omega$ only regulates which PMNS entry talks to which
  HNL, while $\Im(\omega)$ determines the absolute scale of the mixing angle.

  \begin{figure}[!t]
    \centering
    \begin{minipage}{.7\textwidth}
      \centering
      \subfloat[][$\omega =0$\label{Fig. 2HNL diffXSect CI 0}]{\includegraphics[scale=.83,valign=c]{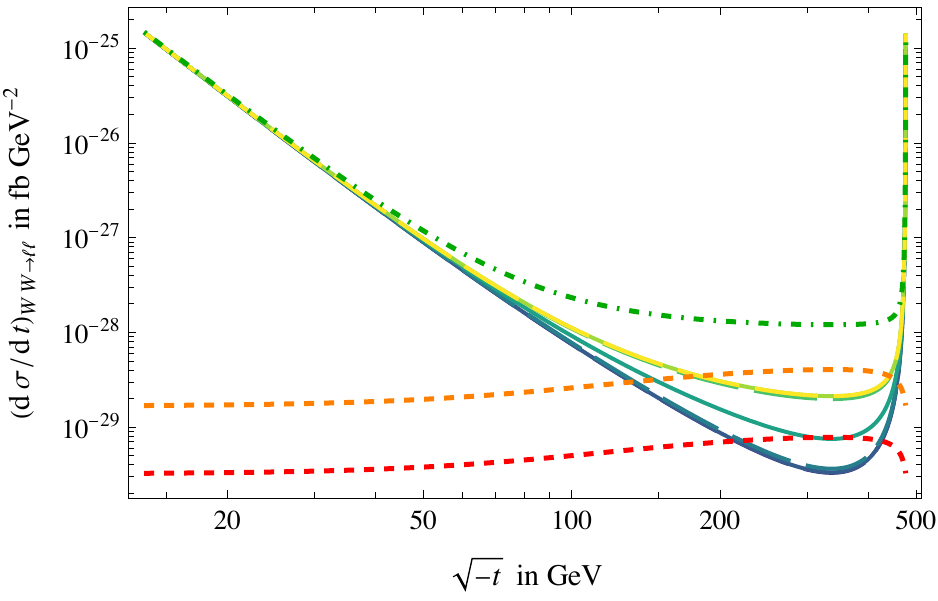}}\\
      \subfloat[][$\omega =5i$\label{Fig. 2HNL diffXSect CI 5I}]{\includegraphics[scale=.83,valign=c]{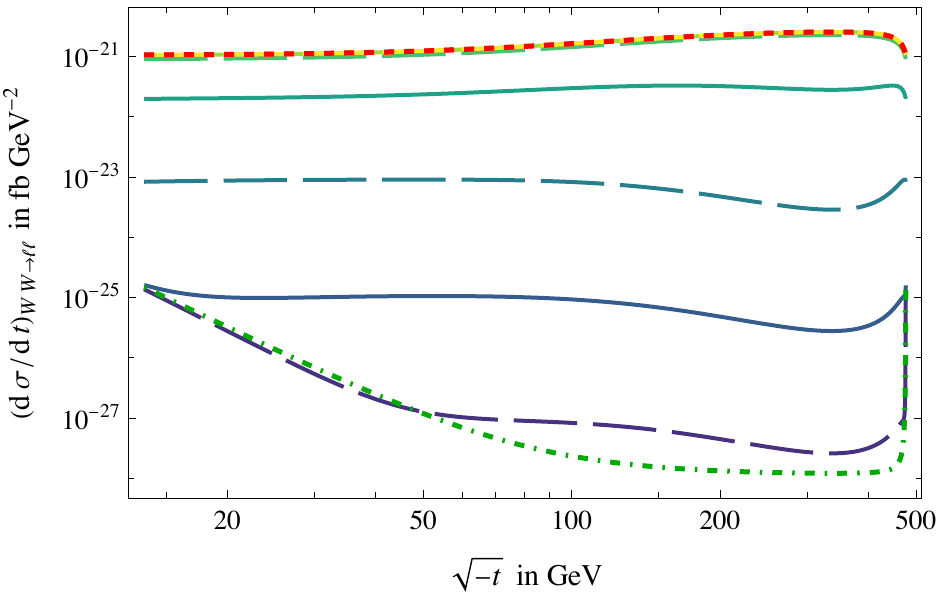}}\\
      \subfloat[][$\omega =15i$\label{Fig. 2HNL diffXSect CI
        15I}]{\includegraphics[scale=.83,valign=c]{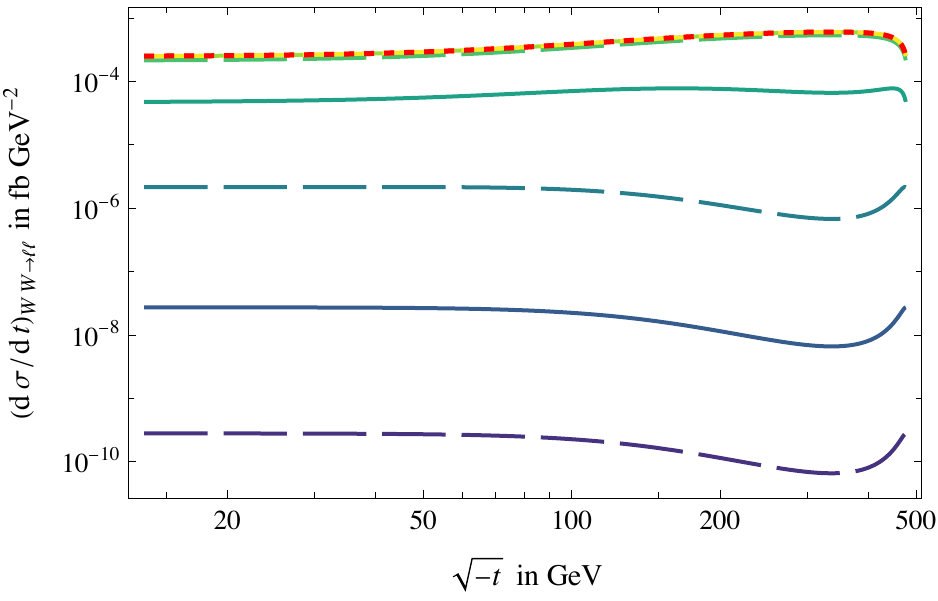}}
    \end{minipage}
    \includegraphics[scale=0.92,valign=c]{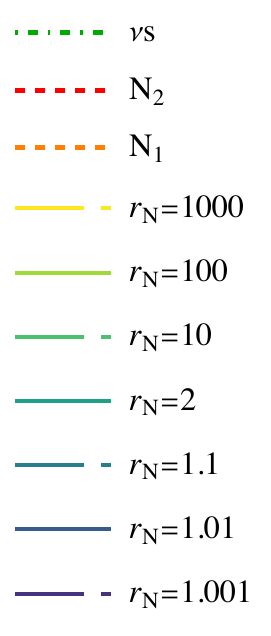}
    \caption{Differential cross section of \WWll scattering using the
      Casas-Ibarra parametrisation of the HNL mixing angles with enhancement
      parameter $\Im(\omega)$.  Here, $r_\Nu=\tfrac{m_{\Nu2}}{m_{\Nu1}}$ is the
      mass ratio.  The lighter of the two HNL masses
      $m_{\Nu_1}=\SI{150}{GeV}$ and the centre of mass energy is
      \SI{490}{GeV}.}
    \label{Fig. 2HNL diffXSect CI}
  \end{figure}

\begin{figure}
    \centering
    \includegraphics[valign=c]{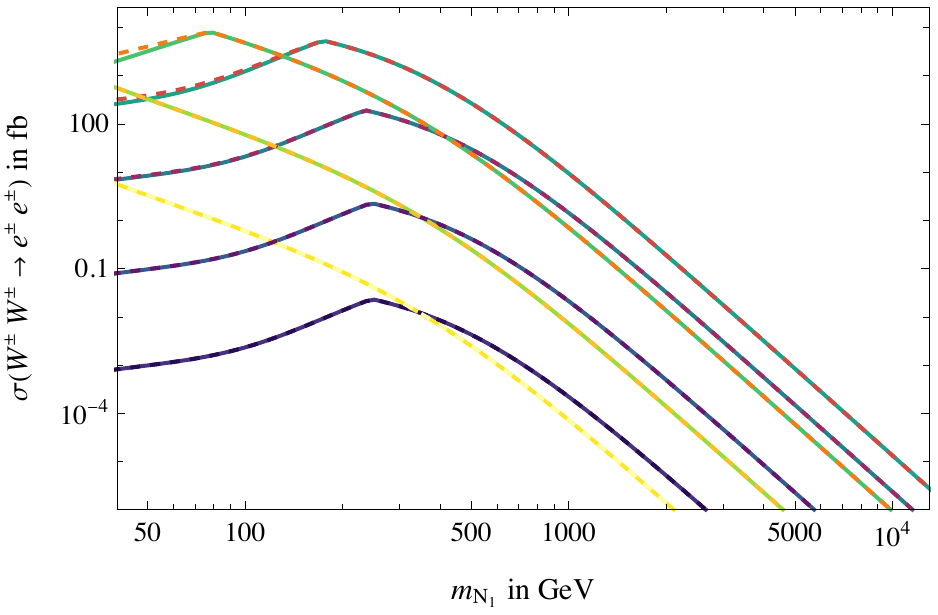}
    \begin{minipage}{.175\textwidth}
      \centering
      {\small \qdl approx.}\\
      \includegraphics[scale=0.92,valign=c]{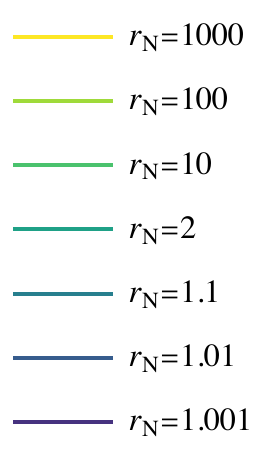}
    \end{minipage}
    \begin{minipage}{.175\textwidth}
      \centering
      {\small Full model}\\
      \includegraphics[scale=0.92,valign=c]{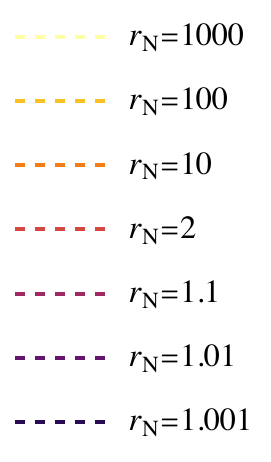}
    \end{minipage}
    \caption{Comparison of \WWll cross sections between \qdl and fully
      realistic two HNL neutrino sector model.  The data is shown at a maximally admissible mixing angle $\mix_{\ell 1}$ according to \cref{eq:Theta_max}. }
    \label{fig:WW 2HNL VMax CIMasslessCompar}
  \end{figure}

  \Cref{Fig. 2HNL diffXSect CI} shows the differential cross section in a model
  with two non-zero light and two heavy mass states for $\omega=0$ (no
  enhancement), $\omega=5i$ (moderate enhancement), and $\omega=15i$ (strong
  enhancement) respectively.  Here, the two HNL curves are both given for a
  mass $m_{\Nu_1}$ but represent the two possible mixing angles of the
  Cassas-Ibarra parametrisation determined by $\Re(\omega)$.  We see that, in
  the case of no enhancement, the situation is rather similar to the single HNL
  case (\cref{Fig. 1HNL diffXSect constLightMass}), where the differential
  cross section is dominated by the light states, and we get a strong
  back-to-back scattering. The strongly enhanced case is essentially the same as
  the quasi-Dirac-like model of two HNLs (\cref{Fig. 2HNL diffXSect Massless}),
  but with a mixing angle slightly smaller than $1$.\footnote{Indeed, it is
    directly comparable with \qdl with a mixing angle of
    $\mix_{\ell 1}^{\mathrm{\qdl}}=|\mix_{\ell 1}^{\mathrm{CI}}(\omega=15i)|<1$
    (see also \cref{fig:WW 2HNL VMax CIMasslessCompar}).}  For large $r_\Nu$,
  the moderately enhanced case exhibits the same characteristics as the
  strongly enhanced case.  However, for smaller $r_\Nu$ the combined
  cross section becomes comparable to that of the isolated light mass states,
  so that non-trivial cancellations occur.

  \Cref{fig:WW 2HNL VMax CIMasslessCompar} compares the \WWll cross section as
  evaluated in a fully realistic 2 HNL model with the same in the \qdl
  approximation. We see that for a maximal theoretically admissible mixing angle
  there is no observable difference between the two results at the scales presented
  here.\footnote{As described in \cref{sec:pp_lljj}, this means that the mixing
  angle fully satisfies either the Seesaw or Yukawa expansion limit.}

  This shows that for large $\mix$, \qdl becomes a very good approximation of a
  fully realistic model.  By virtue of this equivalence, for our purposes, the
  numerical stability of the prior far outweighs the usefulness of
  considering the full model.  This is especially true since only large mixing
  angles have a chance of being detected at the LHC and FCC (see \cref{sec:pp_lljj}).

  \section{Effective \boldmath{W} approximation}
  \label{appx:effectiveWApprox}

\begin{figure}[t]
      \centering
      \subfloat[][$\sqrt{s_\lhc}=\SI{13}{TeV}$\label{Fig. EffectiveW LHC}]{\includegraphics[scale=.7925,valign=c]{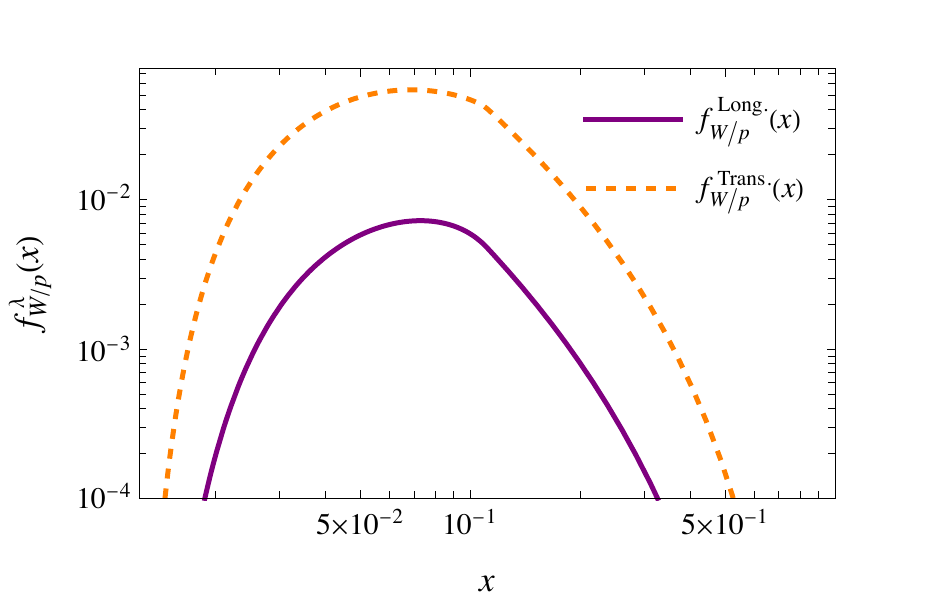}}
      \subfloat[][$\sqrt{s_\fcc}=\SI{100}{TeV}$\label{Fig. EffectiveW FCC}]{\includegraphics[scale=.7925,valign=c]{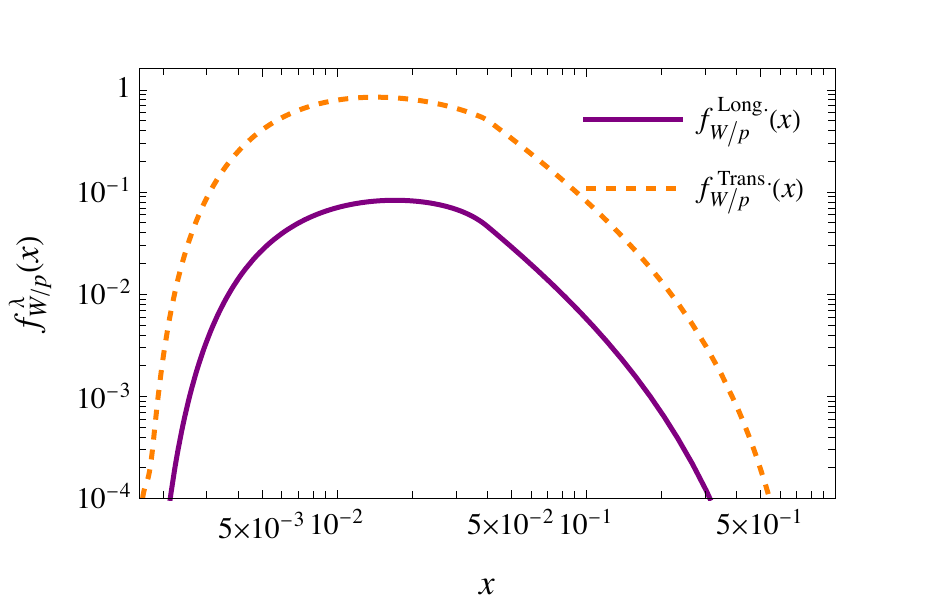}}
      \caption{Parton distribution functions of the effective $W$ approximation as prescribed in \cite{Dawson:1984gx} at LHC \protect\subref{Fig. EffectiveW LHC} and expected FCC centre of mass energy \protect\subref{Fig. EffectiveW FCC}.
      The underlying quark parton distribution functions used were taken from \protect\cite{Martin2009}.}
      \label{fig:EffectiveWPDFs}
    \end{figure}

    In the deduction of the effective $W$ approximation's parton distribution functions (PDFs) according to \cite{Dawson:1984gx} (see \cref{fig:EffectiveWPDFs} for PDFs at relevant centre of mass energies), one formally assumes decoherence between the polarisation states of individual $W$ ``partons".
    This means that if contributions due to different polarisation states are on a similar scale --- and thus the decoherence assumption is no longer valid --- the results at $pp$ level are potentially underestimated by a factor of $\few^2$.\footnote{This is due to the error occurring in both $p$ involved in the \ppll process.}

    \begin{figure}[!t]
      \centering
      \begin{minipage}{.816\textwidth}
        \subfloat[$\lambda_1,\lambda_2=\mathrm{L,L};\,s_\ww=16m_W^2$\label{fig:WWDecompLL16mW2}]{\includegraphics[scale=.81,valign=c]{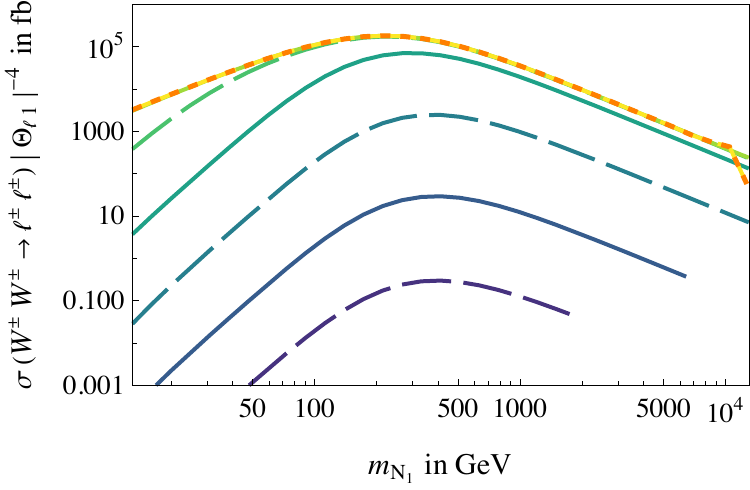}}
        \subfloat[$\lambda_1,\lambda_2=\mathrm{T,T};\,s_\ww=16m_W^2$\label{fig:WWDecompTT16mW2}]{\includegraphics[scale=.81,valign=c]{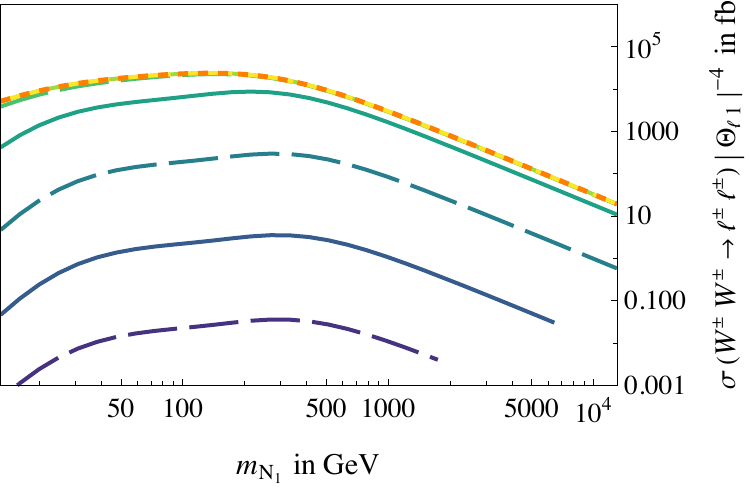}}\\  \subfloat[$\lambda_1,\lambda_2=\mathrm{L,L};\,s_\ww=s_\lhc/100$\label{fig:WWDecompLLSLHCdiv100}]{\includegraphics[scale=.81,valign=c]{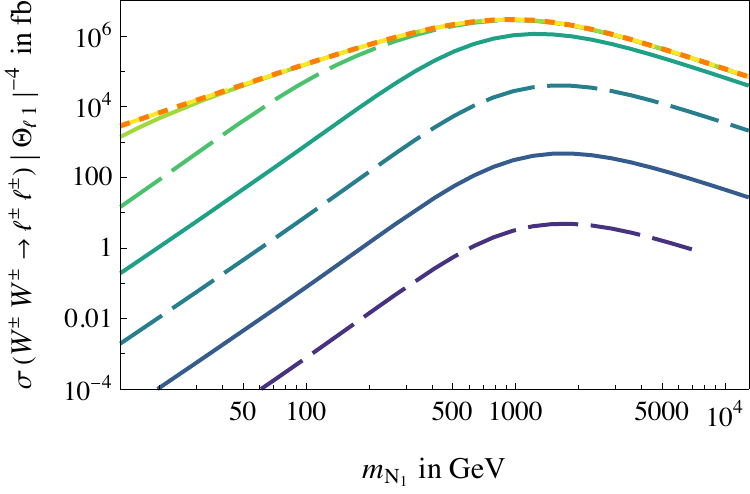}}
          \subfloat[$\lambda_1,\lambda_2=\mathrm{T,T};\,s_\ww=s_\lhc/100$\label{fig:WWDecompTTSLHCdiv100}]{\includegraphics[scale=.81,valign=c]{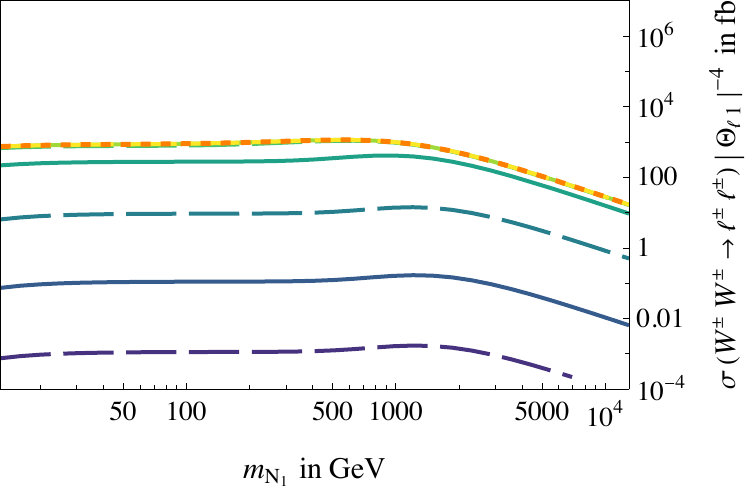}}
      \end{minipage}
      \includegraphics[scale=0.92,valign=c]{Figures/Plots/Legends/2HNL_Degeneracy1NMod.pdf}
      \caption{Polarised cross sections of \WWll.
      The polarisatinos $\lambda_1,\,\lambda_2$ are both longitudinal ($\mathrm{L,L}$, left panels) or transversal ($\mathrm{T,T}$, right panels),
      while the centre of mass energy is $4m_W=\SI{225}{GeV}$ (top panels) or $\sqrt{s_\lhc}/10=\SI{1.3}{TeV}$ (bottom panels).
      We assume a \qdl model with mass ratio $r_\Nu$.}
      \label{fig:WWXSectPolarisationDecomposition}
    \end{figure}

    To investigate the validity of the approximation we will thus compare the doubly longitudinally polarised ($\mathrm{LL}$) $WW$ cross section to the doubly transversally polarised ($\mathrm{TT}$) case.
    As the folding with PDFs involves an integration over different energy regimes, this is important at all relevant centre of mass energies determined by the PDFs in \cref{fig:EffectiveWPDFs}.
    \Cref{fig:WWXSectPolarisationDecomposition} shows this for for a centre of mass energy on the lower end of the PDF spectrum ($s_\ww=16m_W^2$, \cref{fig:WWDecompLL16mW2,,fig:WWDecompTT16mW2}) and on the upper end ($s_\ww=s_\lhc/100$, \cref{fig:WWDecompLLSLHCdiv100,,fig:WWDecompTTSLHCdiv100}).
    We note that for low end energies, the $\mathrm{LL}$ cross section becomes dominant for $m_{\Nu_1}\gtrsim\SIrange{100}{200}{GeV}$, while at larger energies this is the case for $m_{\Nu_1}\gtrsim\SIrange{10}{100}{GeV}$.
    As $pp$ level validity is determined by validity at all relevant subenergies, only a region of general $\mathrm{LL}$ dominance fully meets the assumptions of the effective $W$ approximation.
    We thus expect accurate results in a parameter space of $m_{\Nu_1}\gtrsim\SI{400}{GeV}$, while we expect to underestimate the $pp$ level cross section for smaller $m_{\Nu_1}$.

    \begin{figure}[t]
      \centering
      \includegraphics[scale=.82,valign=c]{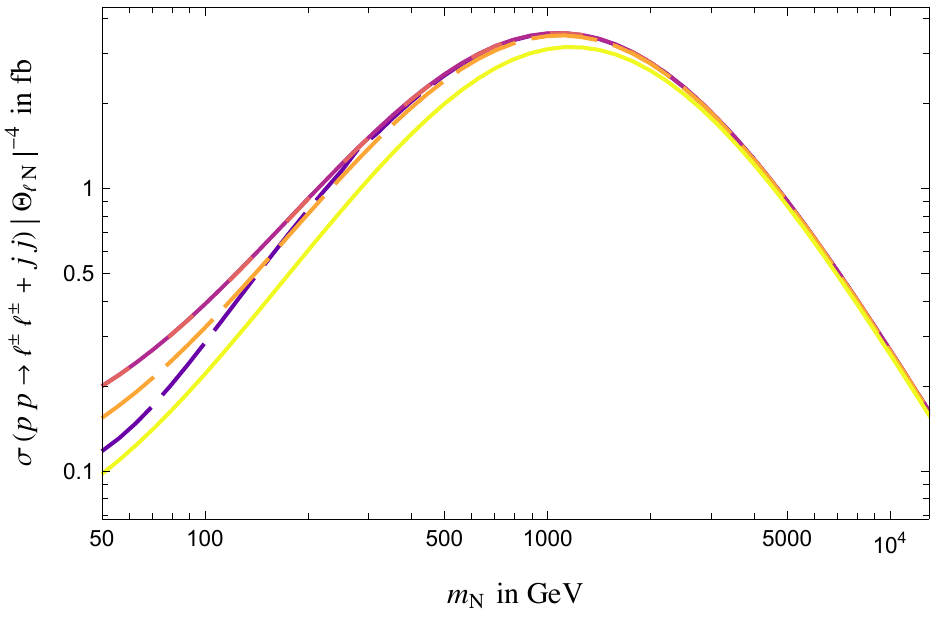}
      \includegraphics[scale=.82,valign=c]{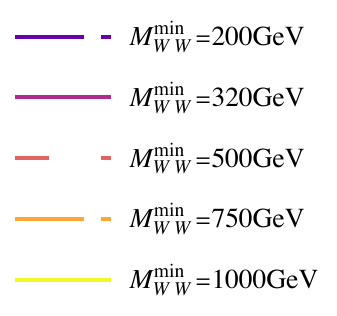}
      \caption{Single isolated HNL $pp$ level WBF cross section at $\sqrt{s_\lhc}=\SI{13}{TeV}$ as a function of HNL mass $m_\Nu$.
      This is shown for different lower bounds on the integration parameters of the effective $W$ approximation determined by $M_{WW}$ according to \cref{eq:x1x2lowlimrelation}.}
      \label{fig:MwwMinCompar}
    \end{figure}

    It has been, furthermore, pointed out in past literature that in order to achieve numerical stability, the lower limit of the integration \cite{Ruiz:2021tdt} has to be set above a certain threshold, rather than to account for minimal production energy.
    We investigate this by varying the lower limit on
    \begin{equation}\label{eq:x1x2lowlimrelation}
      x_1x_2=\frac{M_{WW}^2}{s_{pp}}.
    \end{equation}
    This is shown in \cref{fig:MwwMinCompar}, where we have plotted the cross section of a single isolated HNL for different values of $M_{WW}^2$ at $\sqrt{s_\lhc}=\SI{13}{TeV}$.
    For small HNL masses $m_\Nu\lesssim\SI{100}{GeV}$, there is a noticeable dependence on the lower cut-off.
    We note that the cross section for $M_{WW}^{\min}=\SIrange{320}{500}{GeV}$ is largest in this regime, even though we removed part of the integration interval for positive definite functions with respect to e.g $M_{WW}^{\min}=\SI{200}{GeV}$.
    The lines of $M_{WW}^{\min}=\SI{320}{GeV}$ and $M_{WW}^{\min}=\SI{500}{GeV}$ coincide for all mass values $m_{\Nu_1}$.
    For $m_\Nu\gtrsim\mathrm{several}\,\SI{100}{GeV}$, the numerical values are identical for all $M_{WW}^{\min}<\SI{750}{GeV}$, while the curve of $M_{WW}^{\min}=\SI{750}{GeV}$ lies slightly below the others.
    The curve representing $M_{WW}^{\min}=\SI{1}{TeV}$ lies below the others by a factor of $\mathcal O (\SI{10}{\%})$ for lower $m_\Nu$ converging with the rest at $m_\Nu\simeq\few\,\si{TeV}$.

    Overall the effect of the cut-off is only significant in the regime of $m_\Nu\lesssim\SI{100}{GeV}$.
    As the focus of this work lies on the regime of $m_\Nu\simeq\SIrange[parse-numbers=false]{\mathrm{several}\, 100}{\few\,1000}{GeV}$, the influence is thus marginal.
    Nevertheless, we set the lower limit $M_{WW}^{\min}=4 m_W\simeq\SI{320}{GeV}$ as a benchmark for this work, which corresponds to twice the minimal production energy and is already in a numerically stable regime according to \cref{fig:MwwMinCompar}.

    \begin{figure}
      \centering
      \includegraphics[scale=.92]{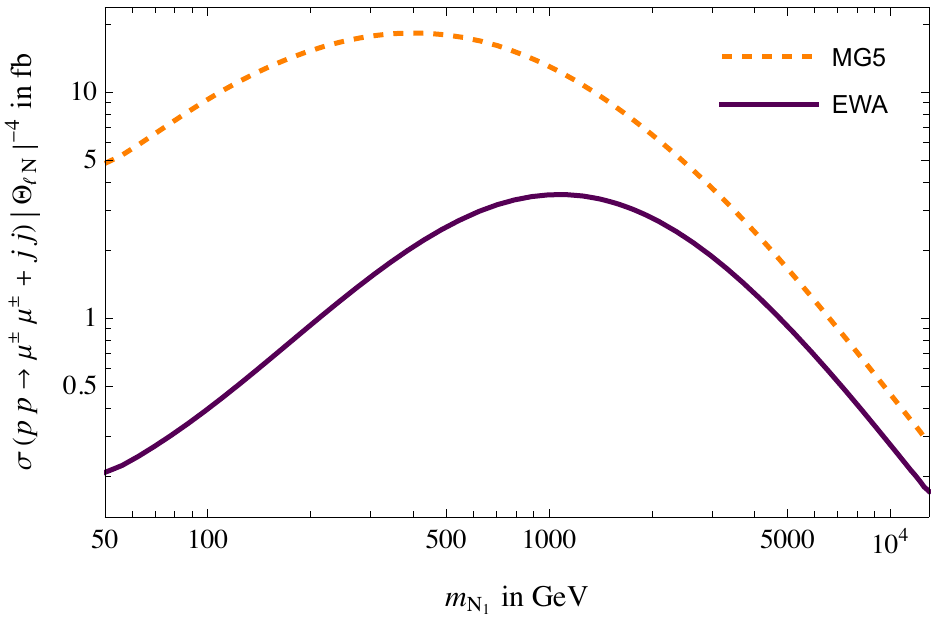}
      \caption{WBF at $pp$ level cross section for a single isolated HNL ($W^\pm W^\pm$ combined).
      Comparison of \madgraph Montecarlo detector simulation by Fuks et al. \cite{Fuks:2020att} (dashed orange) and effective $W$ approximation (purple), shown at LHC centre of mass energy $\sqrt{s_\lhc}=\SI{13}{TeV}$.}
      \label{fig:UsVsFuksXSect}
    \end{figure}

    For the case of a single HNL in a phenomenological \typeI, \cref{fig:UsVsFuksXSect} compares  the resulting cross section deduced using the EWA without any detector-motivated cuts with that given in the NLO in QCD WBF study by Fuks et al. \cite{Fuks:2020att}.
    As expected from the above discussion, the results differ significantly (by a factor of $\sim10$) for $m_{\Nu_1}\sim\few\,\SI{100}{GeV}$.
    For masses in the $\si{TeV}$-range, however, the results become comparable within order $\few$ reaching order $\mathrm{several}\,\SI{10}{\%}$ for $m_{\Nu_1}\sim\mathrm{several}\,\si{TeV}$.
    As this coincides with the significant mass range of our study, the $pp$ level results presented here can be understood as estimates within $\mathcal{O}(1)$.
    We highlight that this only affects the absolute scale of our $pp$ level results, while relative differences between different mass ratios $r_\Nu$ stem from analytic treatment at $WW$ level, and will thus hold at $pp$ level as well.

  \section{Potential avenues for future same sign di-lepton searches at colliders}
  \label{appx:FutureColliderNuBB}

    Even though WBF does not look like a promising channel for HNL discovery at
    the LHC at tree level, it remains to be seen if loop-order corrections to the
    \WWll system could potentially boost the signal.  Especially penguin
    corrections to the $W\Nu\ell$ vertices (similar to \cite{Calderon:2022alb})
    and corrections to the Majorana propagator (similar to those responsible for
    running of light neutrino masses) could be of interest here.

    Another important aspect to investigate is the question of how a tendency toward front-to-back scattering in the $WW$ system for smaller mass ratios translates into angular distributions of the full $pp$ scattering.
    This will need to be investigated in a full on $2\to 4$ simulation of the process as this phenomenon only occurs for $m_{\Nu_1}<\few\,\SI{100}{GeV}$, which is outside the full validity of the EWA, while all angular dependence is strongly suppressed for HNLs with $m_{\Nu_1}\gtrsim\few\,\si{TeV}$.
    If indeed front-to-back scattering translates into large lepton pseudorapidities at $pp$ level, the drop off for smaller $r_\Nu$ with respect to realistic detection prospects could be even more severe than anticipated.
    However, due to the detection limits at the (HL-)LHC, this question is more relevant for FCC energy level searches.

    Another avenue of exploration could be to consider three or more HNLs with similar masses and mixing angles to see if significant LNV effects could be realised in a realistic neutrino sector model.
    As this greatly opens up the available parameter space, \emph{the single HNL case would not hold as an upper limit}.

\end{appendices}

\end{document}